\documentclass[12pt,aps,superscriptaddress,prl]{revtex4-1}

\usepackage{graphicx}  
\usepackage{dcolumn}   
\usepackage{bm}        
\usepackage{amssymb}   
\usepackage{amsmath}   
\usepackage{amsthm}    
\usepackage{xcolor}    
\usepackage{tikz}      
\usepackage{ifthen}    
\usepackage{multirow}  
\usepackage{tabu}      
\usepackage{subfigure}
\usepackage{bbold}
\usepackage{xspace}
\usepackage[unicode=true,
  linktocpage,
  linkbordercolor={0.5 0.5 1},
  citebordercolor={0.5 1 0.5},
  linkcolor=blue]{hyperref}
\newcommand{\sro}{Sr$_2$RuO$_4$\xspace}
\newcommand{\Tc}{$T_\mathrm{c}$\xspace}
\newcommand{\highTc}{high-$T_\mathrm{c}$\xspace}

\newcounter{para}

\begin{document}
	
\title{Thermodynamic Evidence for a Two-Component Superconducting Order Parameter in Sr$_2$RuO$_4$}

\author{Sayak Ghosh}
\affiliation{Laboratory of Atomic and Solid State Physics, Cornell University, Ithaca, NY 14853, USA}
\author{Arkady Shekhter}
\affiliation{National High Magnetic Field Laboratory, Florida State University, Tallahassee, FL 32310, USA}
\author{F. Jerzembeck}
\affiliation{Max Planck Institute for Chemical Physics of Solids, Dresden, Germany}
\author{N. Kikugawa}
\affiliation{National Institute for Materials Science, Tsukuba, Ibaraki 305-0003, Japan}
\author{Dmitry A. Sokolov}
\affiliation{Max Planck Institute for Chemical Physics of Solids, Dresden, Germany}
\author{Manuel Brando}
\affiliation{Max Planck Institute for Chemical Physics of Solids, Dresden, Germany}
\author{A. P. Mackenzie}
\affiliation{Max Planck Institute for Chemical Physics of Solids, Dresden, Germany}
\author{Clifford W. Hicks}
\affiliation{Max Planck Institute for Chemical Physics of Solids, Dresden, Germany}
\author{B.~J.~Ramshaw$^\ast$}
\affiliation{Laboratory of Atomic and Solid State Physics, Cornell University, Ithaca, NY 14853, USA}

\maketitle

\textbf{\sro has stood as the leading candidate for a spin-triplet superconductor for 26 years. Recent NMR experiments have cast doubt on this candidacy, however, and it is difficult to find a theory of superconductivity that is consistent with all experiments. What is needed are symmetry-based experiments that can rule out broad classes of possible superconducting order parameters. Here we use resonant ultrasound spectroscopy to measure the entire symmetry-resolved elastic tensor of \sro through the superconducting transition. We observe a thermodynamic discontinuity in the shear elastic modulus $c_{66}$, requiring that the superconducting order parameter is two-component. A two-component $p$-wave order parameter, such as $p_x+i p_y$, naturally satisfies this requirement. As this order parameter appears to be precluded by recent NMR experiments, we suggest that two other two-component order parameters, namely $\left\{d_{xz},d_{yz}\right\}$ or $\left\{d_{x^2-y^2},g_{xy(x^2-y^2)}\right\}$, are now the prime candidates for the order parameter of \sro.}

\section{Introduction}

Nearly all known superconductors are ``spin-singlet'', composed of Cooper pairs that pair spin-up electrons with spin-down electrons. Noting that \sro has similar normal-state properties to superfluid $^3$He, Rice and Sigrist\cite{RiceIOP1995} and, separately, Baskaran \cite{baskaran1996sr2ruo4}, suggested that \sro may be a solid-state ``spin-triplet'' superconductor. This attracted the attention of the experimental community, and ensured decades of intense research on \sro that resulted in a detailed understanding of its metallic state \cite{maeno1994superconductivity,BergemannPRL2000}. From this well-understood starting point one might expect the superconductivity of \sro to be a solved problem \cite{MackenzieRMP2003}, but decades after its discovery the symmetry of the superconducting order parameter remains a mystery, largely due to discrepancies between several major pieces of experimental evidence \cite{MackenzieNPJ2017}. 

Formerly, the strongest evidence for spin-triplet pairing in \sro was a Knight shift that was unchanged upon entering the superconducting state \cite{IshidaNature1998}. A recently revised version of this experiment, however, shows that the Knight shift is suppressed below \Tc, ruling against most spin-triplet order parameters \cite{PustogowNat2019,ishida2019reduction}. This is consistent with measurements of the upper-critical magnetic field, which appears to be Pauli-limited and thus suggests spin-singlet pairing \cite{Kittaka:2009}. The challenge is to reconcile these data with previous evidence in favour of a spin-triplet order parameter, including time-reversal symmetry breaking below \Tc in $\mu$SR\cite{LukeNature1998} and polar Kerr effect experiments\cite{XiaPRL2006}, and half-quantized vortices \cite{JangScience2011}.   

While the spin-triplet versus spin-singlet aspect of the superconductivity in \sro is still under debate, less well studied is the symmetry of the orbital part of the Cooper pair wavefunction. By symmetry, spin-triplet superconductors are required to have an odd-parity orbital wavefunction, i.e. to be an $l=1$ `$p$-wave' or $l=3$ `$f$-wave' superconductor, where $l$ is the orbital quantum number. This is in contrast with conventional $l=0$ $s$-wave superconductors, or the \highTc $l=2$ $d$-wave superconductors. While some information about the orbital wavefunction can be inferred by looking for nodes in the superconducting gap, determination of nodal position does not uniquely determine the orbital structure of the Cooper pair. 

\begin{figure*}[h!]
	\centering
	\includegraphics[width=.99\linewidth]{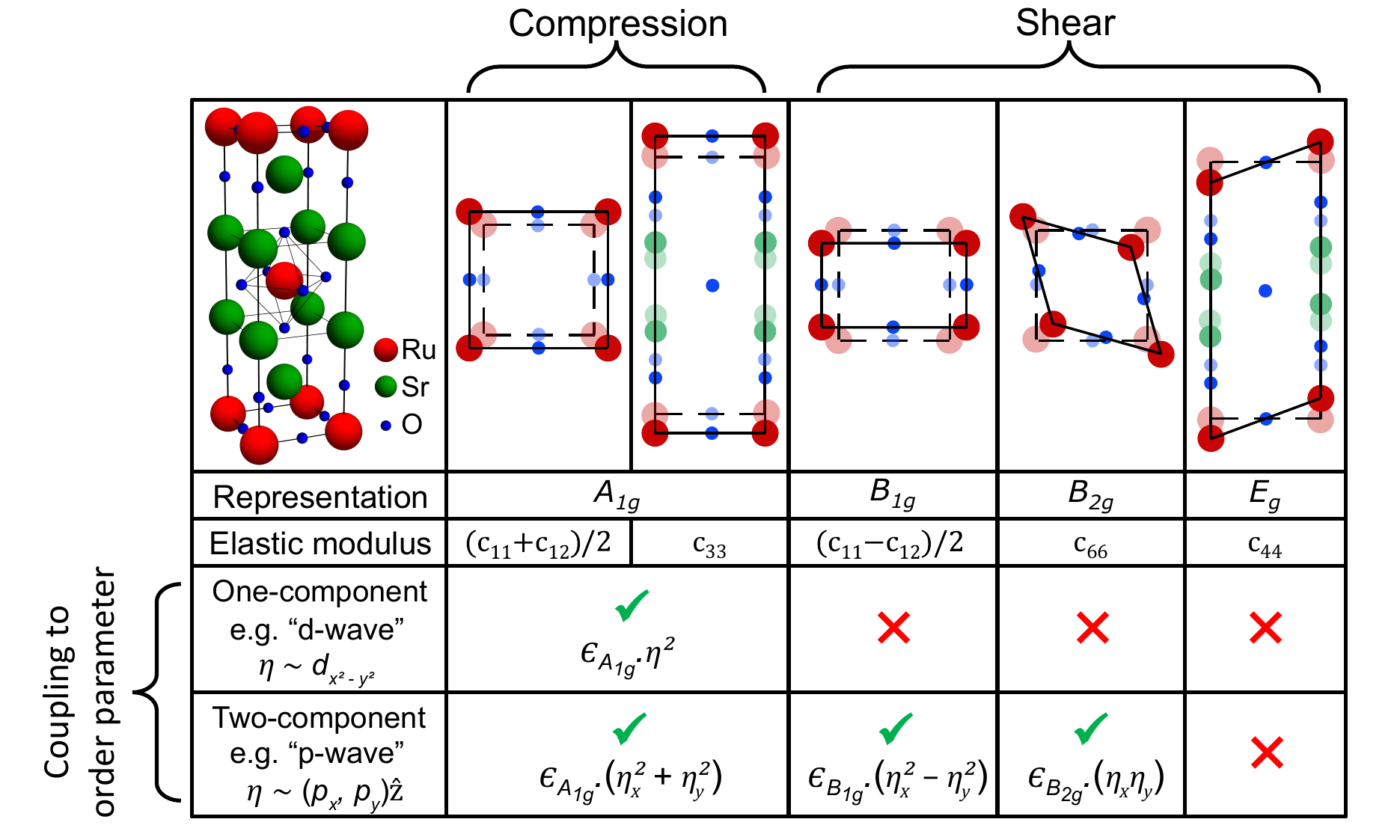}
	\caption{\textbf{Irreducible strains in \sro and their coupling to superconducting order parameters.} The tetragonal crystal structure of \sro and unit cell deformations illustrating the irreducible representations of strain are shown. There is an elastic modulus corresponding to each of these strains, and a sixth modulus $c_{13}$ that arises from coupling between the two $A_{1g}$ strains. Green check marks denote allowed linear-order couplings to strain for one and two-component order parameter bilinears, and red crosses denote that such coupling is forbidden. These couplings are what lead to discontinuities in the elastic moduli at \Tc. See \autoref{tab:OPtable} for a list of relevant possible order parameters in \sro. }
	\label{fig:coupling}
\end{figure*}

One way to distinguish different orbital states is by their degeneracy---the number of states with the same energy. $s$-wave and $d_{x^2-y^2}$-wave Cooper pairing states, for example, are both singly degenerate (``one-component''), while the $\left\{p_x,p_y\right\}$ state (which can order in the chiral $p_x+ip_y$ configuration) is two-fold degenerate (``two-component''). This difference in orbital degeneracy has an unambiguous signature in an ultrasound experiment: shear elastic moduli are continuous through \Tc for a singly-degenerate orbital state, but are discontinuous across \Tc for a doubly degenerate state \cite{RehwaldAIP1973,GhoshSciAdv2020}. The observation of a discontinuity in one of the shear elastic moduli of \sro at \Tc would therefore constitute strong evidence in favour of either $p$-wave superconductivity, or one of the other two-component superconducting order parameters. These measurements have been attempted in the past, and were suggestive of a shear discontinuity at \Tc, but a discontinuity was also found in a symmetry-forbidden channel, and the experiment was thus deemed to be inconclusive \cite{okuda2002unconventional}. Other independent evidence of a shear discontinuity \cite{lupien2002ultrasound} is now being submitted as part of a separate complementary study using an experimental technique different from our own (for a theoretical interpretation of these results, see \citet{WalkerPRB2002}). 

\section{Experiment}


Elastic moduli are second derivatives with respect to strain of a system's total free energy. Elastic moduli are therefore thermodynamic coefficients akin to heat capacity or magnetic susceptibility, and are indicative of a system's ground-state properties. Strain is a second-rank tensor quantity, and thus it can couple to order parameters in ways that lower-rank quantities, such as temperature and electric field, cannot. This, in particular, requires that elastic moduli behave differently in systems with one- or two-component order parameters. Here we provide a brief overview of the connection between crystal symmetry, order parameter symmetry, and ultrasound: the detailed derivations can be found in the `Strain-Order Parameter Coupling' section in S.I., as well as in a number of theoretical papers \cite{RehwaldAIP1973,WalkerPRB2002,SigristPTP2002}. 

\begin{figure*}[h!]
	\centering
	\includegraphics[width=.99\linewidth]{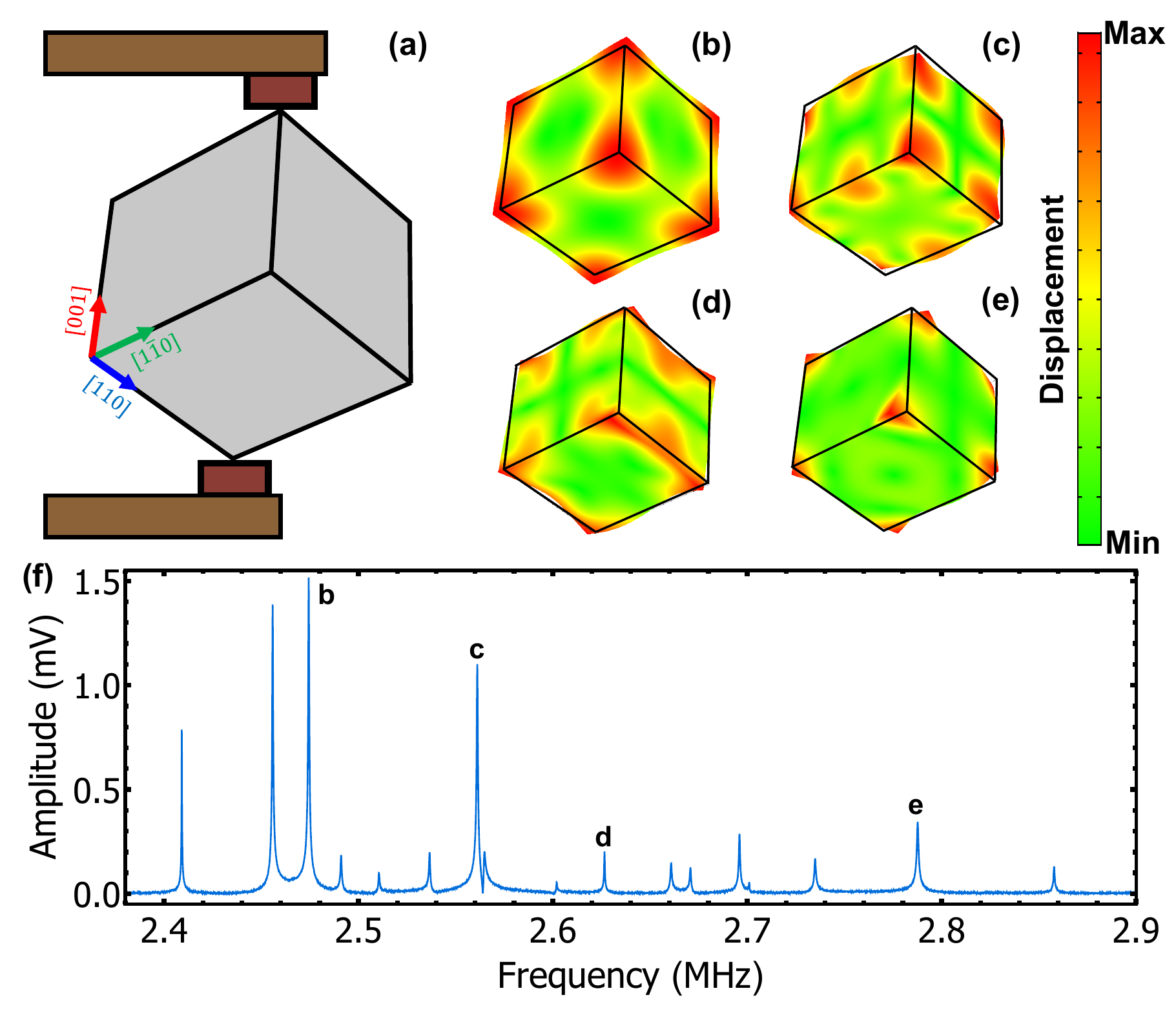}
	\caption{\textbf{Resonant ultrasound spectroscopy: schematic and spectrum.} (a) A single-crystal sample, polished along known crystal axes, is held in weak-coupling contact between two ultrasonic transducers, allowing it to vibrate freely at its resonance frequencies. Panels (b) through (e) show the crystal's deformation corresponding to four particular experimentally measured resonances, marked in (f). (f) A portion of the ultrasonic spectrum of \sro, in the frequency range from 2.4-2.9 MHz, taken at room temperature. Each resonance creates a unique strain pattern in the material that can be decomposed in terms of the five irreducible strains (\autoref{fig:coupling}(a)), modulated in phase along the dimensions of the sample.}
	\label{fig:RUS}
\end{figure*}

The allowed couplings between strains and superconducting order parameters become transparent when both are described in terms of irreducible representations (irreps) of the point-group symmetry.  \sro crystallizes in the tetragonal space group $I4/mmm$, along with its associated point group $D_{4h}$. In this crystal field environment, the five-component $l=2$ $d$-representation breaks into three one-component irreps: $d_{z^2}$ ($A_{1g}$ irrep), $d_{xy}$ ($B_{2g}$), and $d_{x^2-y^2}$ ($B_{1g}$---the familiar `$d$-wave' of the cuprates), and one two-component irrep $\left\{d_{xz},d_{yz}\right\}$ ($E_g$). The three-component $p$-representation breaks into the one-component irrep $p_z$ ($A_{2u}$) and the two-component irrep $\left\{p_x,p_y\right\}$ ($E_{u}$)---it is this latter representation that has been proposed to order into the chiral $p_x + ip_y$ superconducting state.										

As illustrated in \autoref{fig:coupling}, there are five unique strains ($\epsilon_{\Gamma}$) in \sro (five irreps ($\Gamma$) of strain in $D_{4h}$): two compressive strains transforming as the $A_{1g}$ irrep, and three shear strains transforming as the $B_{1g}$, $B_{2g}$ and $E_{g}$ irrep. Each strain has a corresponding elastic modulus, $c_{\Gamma}=\partial^2\mathcal{F}/\partial\epsilon_{\Gamma}^2$, where $\mathcal{F}$ is the thermodynamic free energy. A sixth modulus, $c_{13}$, defines the coupling between the two $A_{1g}$ strains ($\epsilon_{xx} + \epsilon_{yy}$ and $\epsilon_{zz}$.) Sound velocities can be computed from these moduli as $v_{\Gamma} = \sqrt{c_{\Gamma}/\rho}$, where $\rho$ is the density. When composing terms in the free energy, direct (linear) coupling between strain and the superconducting order parameter ($\eta$) is forbidden because superconductivity breaks gauge symmetry. The next relevant coupling is linear in strain and quadratic in order parameter. For one-component superconducting order parameters, including all $s$-wave states and the $d_{x^2-y^2}$ state, the only quadratic form is $\eta^2$, transforming as $A_{1g}$, and thus the only allowed coupling is $\epsilon_{A_{1g}}\eta^2$. This coupling produces discontinuities in all the $A_{1g}$ (compressional) elastic moduli across \Tc. Two-component order parameters ($\vec{\eta} = \left\{\eta_x,\eta_y\right\}$), on the other hand, have three independent quadratic forms: $\eta_x^2+\eta_y^2$, $\eta_x^2-\eta_y^2$, and $\eta_x\eta_y$, transforming as $A_{1g}$, $B_{1g}$, and $B_{2g}$, respectively. Thus in addition to coupling to the $A_{1g}$ elastic moduli, two-component order parameters couple to two of the shear moduli through $\epsilon_{B_{1g}}\left(\eta_x^2-\eta_y^2\right)$ and $\epsilon_{B_{2g}} \eta_x \eta_y$. This produces discontinuities in the associated shear elastic moduli ($\left(c_{11}-c_{12}\right)/2$ and $c_{66}$, respectively), based purely on symmetry considerations, and independent of the microscopic mechanism of superconductivity. 

\begin{figure*}[h!]
	\centering
	\includegraphics[width=.99\linewidth]{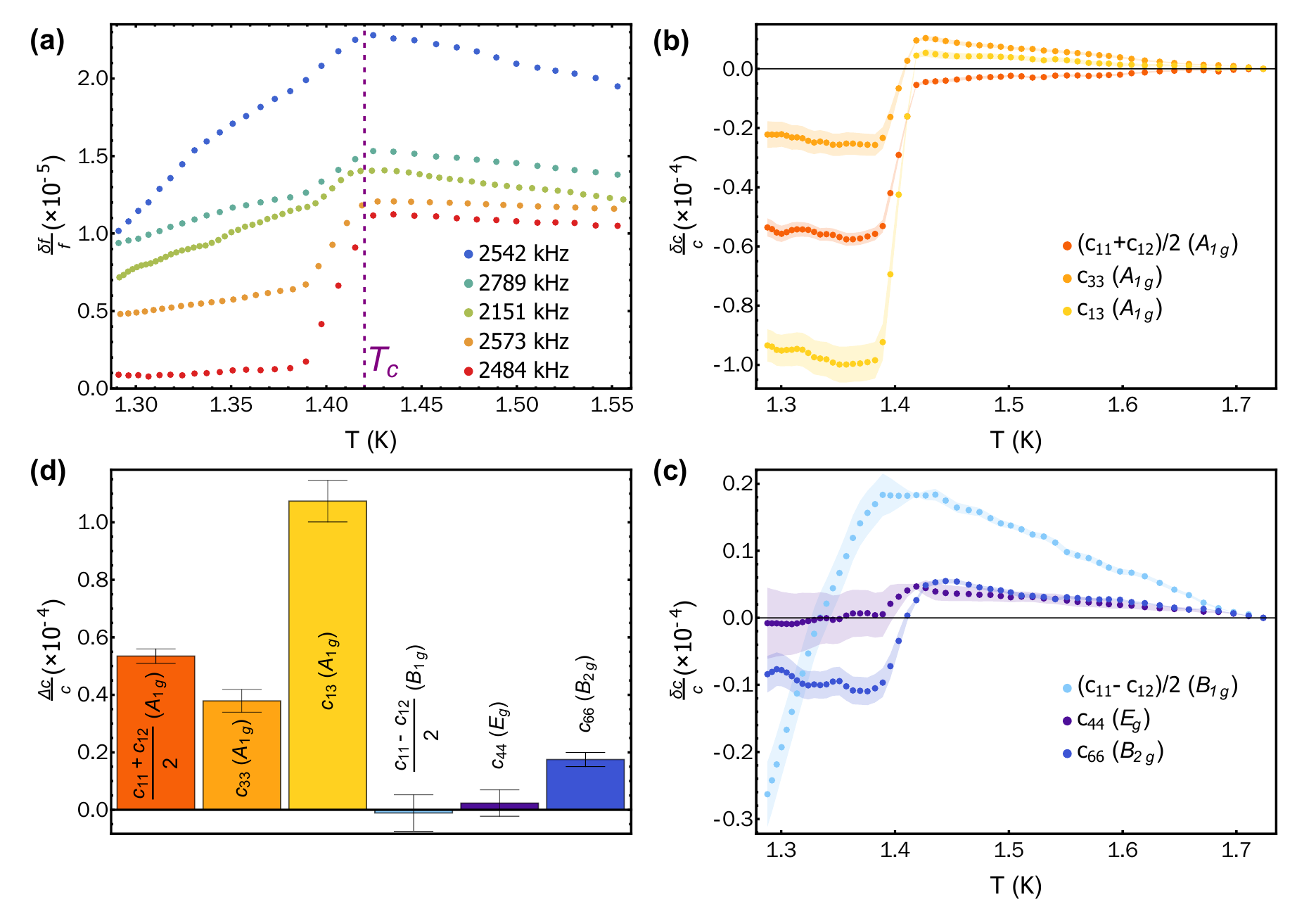}
	\caption{\textbf{Resonant ultrasound spectroscopy across \Tc in \sro.} (a) Temperature evolution of five representative resonances measured through \Tc  --- plots are shifted vertically for visual clarity. The dashed line shows \Tc determined by resistivity measurements. A step-like discontinuity (``jump'') is observed at \Tc---the different magnitudes of this jump signify the contributions of different elastic moduli in each resonance. 18 such resonances were tracked through \Tc to determine the elastic moduli. (b) Compressional ($A_{1g}$) and (c) shear ($B_{1g}$, $E_{g}$ and $B_{2g}$) moduli of \sro across \Tc, along with their experimental errors arising from the uncertainties in sample dimensions. The absolute values of these moduli at 4 kelvin are determined to be $(c_{11}+c_{12})/2 = 190.8$, $c_{33}=257.2$, $c_{13}=85.0$, $(c_{11}-c_{12})/2 =53.1$, $c_{66}=65.5 $ and $c_{44}=69.5$  GPa. (d) Magnitudes of the elastic moduli jumps at \Tc, along with their experimental uncertainties (see `Uncertainty Analysis' in S.I. for details).}
	\label{fig:moduli}
\end{figure*}

While traditional pulse-echo ultrasound experiments measure a single elastic modulus per experiment, we use resonant ultrasound spectroscopy (RUS) to measure all six elastic moduli of \sro across \Tc in a single experiment, greatly reducing systematic uncertainty \cite{migliori1993resonant}. Analogous to how a stretched string has standing waves that can be expressed in terms of sinusoidal harmonics, three-dimensional solids have elastic resonances that can be decomposed in terms of the irreps of strain. RUS measures the frequencies of these resonances for a single crystal sample, from which all elastic moduli can be obtained by inverse-solving the elastic wave equation \cite{RamshawPNAS}. The relatively low \Tc($\approx$ 1.43 K for our sample) of \sro, however, poses severe technical challenges to perform RUS experiments across \Tc. RUS samples are typically large (of the order 1 mm$^3$), and are only in weak-coupling contact with the transducers (see \autoref{fig:RUS}(a)). This ensures nearly-free boundary conditions but prevents good thermal coupling between the apparatus and the sample. Previous RUS implementations either sacrificed uniform cooling by placing the sample in vacuum \cite{RamshawPNAS}, or sacrificed a slow cooling rate by placing the sample in direct thermal contact with the helium bath. Our new RUS design employs a double vacuum can arrangement (see Methods for details) to allow for slow, uniform cooling of a sample down to approximately 1.25 K. We observe a sharp (40 mK wide) superconducting transition (\autoref{fig:moduli}(a)), signifying high sample quality and uniform sample cooling.  

We performed RUS measurements on a single-crystal sample of \sro across \Tc. The full data set includes 18 resonances (see \autoref{fig:fvT})---five representative resonances are shown in \autoref{fig:moduli}(a). We decompose all 18 resonances into the three compressional moduli ($\left(c_{11}+c_{12}\right)/2$, $c_{33}$, and $c_{13}$) and the three shear moduli ($\left(c_{11}-c_{12}\right)/2$, $c_{44}$, and $c_{66}$.) The temperature evolution of these moduli are shown in \autoref{fig:moduli}(b) and (c). Discontinuities across \Tc are clearly observed in all three compressional moduli, as required by thermodynamics for all superconductors, as well as in the shear modulus $c_{66}$. The discontinuity in $c_{66}$ is forbidden by symmetry for one-component order parameters, but is allowed for two-component order parameters---this discontinuity is our central finding. 

Having measured all six elastic moduli across \Tc, we can perform several consistency checks. First, since there is no bilinear coupling of the order parameter to $E_{g}$ strain, $c_{44}$ should not have a discontinuity at \Tc for any superconducting order parameter. Within our experimental uncertainty, we observe only a change in the slope of $c_{44}$ at \Tc, which is expected on general grounds and is not constrained by symmetry \cite{RehwaldAIP1973}. Second, thermodynamics dictates that the discontinuities in the three compressional moduli at a second order phase transition should follow a self-consistency relation (see discussion preceding Equation S16 in S.I.),
\begin{equation}
 \Big(\Delta \frac{c_{11} +c_{12}}{2}\Big) \times (\Delta c_{33}) = (\Delta c_{13})^2.
\label{eq:jump}
\end{equation} 
From our measurement, we find $(\Delta \frac{c_{11} +c_{12}}{2}) \times (\Delta c_{33}) = (9.9\pm1.5)\!\times\! 10^{-5}~ \mathrm{GPa}^2$ and $(\Delta c_{13})^2 = (8.3\pm1.1)\!\times\! 10^{-5}~ \mathrm{GPa}^2$. These consistency checks validate our measurement of the magnitude of the jumps, and our ability to correctly decompose the jumps in frequency into jumps in the irreducible elastic moduli.

A further check on the data is provided by an Ehrenfest relation that relates the derivative of \Tc with hydrostatic pressure, $P$, to the discontinuities at \Tc in the specific heat, $\Delta C$, and the bulk modulus, $\Delta B$, via 
\begin{equation}
\left(\frac{dT_c}{dP}\right)^2 = -\frac{\Delta B}{B^2}\left(\frac{\Delta C}{T}\right)^{-1}.
\label{eq:eh}
\end{equation}
We measure $\Delta B/B \approx 6.3 \times 10^{-5}$ (see `Ehrenfest Relations for Compressional Strains' in S.I.). Combined with $\Delta C/T$ for this sample (see \autoref{fig:sampleChar}), \autoref{eq:eh} yields a value for $dT_c/dP$ of 0.90 K/GPa. This is a factor of 3 higher than what is reported for a direct measurement of \Tc as a function of pressure \cite{forsythe2002evolution}. This discrepancy may be evidence for a pair of transitions occurring at or near the superconducting \Tc, as discovered by recent $\mu$SR experiments \cite{grinenko2020split}. The two transition temperatures split when stress is applied along the $x$ direction---Meissner screening onsets at the higher transition temperature, $T_c$, while time reversal symmetry is broken at the lower transition, $T$ \mbox{\tiny $\!\!\!\!_{TRSB}$ }. To perform the correct Ehrenfest analysis one would require $dT$ \mbox{\tiny $\!\!\!\!_{TRSB}$ } $\!\!/dP$, which is unknown at present.
A similar Ehrenfest relation---derived for the jump in $c_{66}$ rather than the jump in bulk modulus---requires that \Tc shift linearly with $B_{2g}$ strain, specifically as \Tc$\propto \left|\epsilon_{xy}\right|$. Prior measurements of \Tc as a function of $\epsilon_{xy}$, however, have not found this linear dependence on strain \cite{HicksScience2014}. In addition, most ordered states of two-component order parameters should exhibit two transition temperatures under finite strain, but this has not been found either by heat capacity or local susceptibility measurements \cite{LiArxiv2019,watson2018micron}, at least for $B_{1g}$ strain. In the S.I. we show that the current experimental resolution on these phenomena is still consistent with the size of the jump that we find in $c_{66}$ (see `Reconciling the $c_{66}$ Discontinuity with Experiments Under Finite Strain').																							

\section{Discussion}

\begin{table*}
	\centering	
	\begin{tabular}{| c | c | c | c | c | c |}
		\hline
		Dimensionality & Order Parameter & Irrep. & Moduli Jumps & Ultrasound & NMR  \\ 
		\hline\hline
		
		\multirow{3}{3.5em}{One-component} & $s$ & A$_{1g}$ & A$_{1g}$ & $\times$ &  \checkmark \\
		\cline{2-6}
		& $d_{x^2-y^2}$ & B$_{1g}$ & A$_{1g}$ & $\times$ & \checkmark \\
		\cline{2-6}
		& $d_{xy}$ & B$_{2g}$ & A$_{1g}$ & $\times$ &  \checkmark \\
		\hline
		\multirow{4}{3.5em}{Two-component}
		& $\left\{p_x,p_y\right\}\hat{\boldsymbol{z}}$ & E$_u$ & A$_{1g}$, B$_{1g}$, B$_{2g}$ &  \checkmark & $\times$\\
		\cline{2-6}
		& $p_{z}\left\{\hat{\boldsymbol{x}},\hat{\boldsymbol{y}}\right\}$ & E$_u$ & A$_{1g}$, B$_{1g}$, B$_{2g}$ &  \checkmark & $\times$\\
		\cline{2-6}
		& $\left\{d_{xz}, d_{yz}\right\}$ & E$_g$ & A$_{1g}$, B$_{1g}$, B$_{2g}$ &  \checkmark &  \checkmark \\ 
				\cline{2-6}
		 & $\left\{d_{x^2-y^2},g_{xy(x^2-y^2)}\right\}$ & B$_{1g}$ $\oplus$ A$_{2g}$ & A$_{1g}$, B$_{2g}$&  \checkmark &  \checkmark \\
		\hline	
	\end{tabular}
	\caption{\textbf{Some superconducting order parameters and their representations in $D_{4h}$.} For the odd-parity spin-triplet order parameters, $\hat{\boldsymbol{x}}$, $\hat{\boldsymbol{y}}$ and $\hat{\boldsymbol{z}}$ represent the pair wavefunction in spin space in the $\mathbf{d}$-vector notation \cite{MackenzieRMP2003}. Two component order parameters $\left\{\eta_x,\eta_y\right\}$ can order as $\eta_x$, $\eta_y$, $\eta_x \pm \eta_y$, or $\eta_x \pm i \eta_y$, depending on microscopic details. It is this latter combination that forms the time-reversal symmetry breaking state (e.g. $(p_x + i p_y)\hat{\boldsymbol{z}}$, or $d_{xz} + i d_{yz}$). The ``Ultrasound'' column indicates whether an order parameter is consistent with a jump in $c_{66}$ at \Tc, the ``NMR'' column indicates whether it is consistent with the suppression of the Knight shift at \Tc. Note that the $B_{1g} \oplus A_{2g}$ state does not belong to a single irrep of $D_{4h}$, and thus transition temperatures of the $d$ and $g$ components must be ``fine-tuned" if they are to coincide.}
	\label{tab:OPtable}
\end{table*}

A discontinuity in $c_{66}$ at \Tc can \textit{only} result from a two-component superconducting order parameter (see \autoref{tab:OPtable}). This is a critical piece of information because evidence for vertical line nodes in the superconducting gap---from ultrasonic attenuation \cite{lupien2002ultrasound}, heat capacity \cite{deguchi2004determination}, thermal conductivity \cite{HassingerPRX2017}, and quasiparticle interference \cite{sharma2019momentum}---are most straightforwardly interpreted in terms of a one-component, $d_{x^2-y^2}$, order parameter. With the discovery of a suppression of the Knight shift strongly suggesting that the order parameter cannot be spin-triplet \cite{PustogowNat2019}, $d_{x^2-y^2}$ would seem a likely contender. The discontinuity in $c_{66}$, however, rules against \textit{any} one-component order parameter, including $d_{x^2-y^2}$. 

Our measurement is consistent with several two-component $p$-wave scenarios, including $(p_{x} \pm i p_{y})\hat{\boldsymbol{z}}$ and $p_{z}(\hat{\boldsymbol{x}} \pm i \hat{\boldsymbol{y}})$. Taken at face value, however, the suppression of the Knight shift \cite{PustogowNat2019} rules out all $p-$wave order parameters, and is consistent only with spin-singlet order parameters. The only ``conventional'' spin-singlet order parameter that produces a jump in $c_{66}$ at \Tc is $\left\{d_{xz}, d_{yz}\right\}$. This state can order into the non-magnetic $d_{xz}$, $d_{yz}$, or $d_{xz} \pm d_{yz}$ states, all of which break the $C_4$ rotational symmetry of the lattice. It can also order into the chiral magnetic $d_{xz} \pm i d_{yz}$ state. If one considers the possibility of an accidental degeneracy between two order parameters of different representations, producing accidental two-component order parameters, then $\left\{d_{xy},s\right\}$ and $\left\{d_{x^2-y^2}, g_{xy(x^2-y^2)}\right\}$ are also consistent with our experiment ($\left\{d_{x^2-y^2},s\right\}$ \cite{romer2019knight} produces a jump in $\left(c_{11}-c_{12}\right)/2$ but not in $c_{66}$). We set aside $\left\{d_{xy},s\right\}$ because it thought to be incompatible with the electronic structure of \sro. If one accepts time-reversal symmetry breaking at \Tc as a property of \sro, there are then two remaining order parameters that are compatible with our experiment: $d_{xz} \pm i d_{yz}$, and $d_{x^2-y^2} \pm i g_{xy(x^2-y^2)}$. The absence of a discontinuity in $(c_{11}-c_{12})/2$ implies that there is no order-parameter-bilinear that transforms as B$_{1g}$, which would rule out $d_{xz} \pm i d_{yz}$. It is possible, however, that while a jump in $(c_{11}-c_{12})/2$ is required thermodynamically, it is either unobservably small because the coupling coefficient is small (for microscopic reasons), or the jump is smeared-out due to high ultrasonic attenuation in the $B_{1g}$ channel \cite{LupienPRL2001}. Thus we consider the implications of both the $d_{xz} \pm i d_{yz}$ and the $d_{x^2-y^2} \pm i g_{xy(x^2-y^2)}$ superconducting states in \sro.

The first of these, $d_{xz} \pm i d_{yz}$, is the chiral-ordered state of $\left\{d_{xz}, d_{yz}\right\}$---a two-component $E_g$ representation \cite{zutic2005phase}. There are two main arguments against such a state. First, $\left\{d_{xz}, d_{yz}\right\}$ has a horizontal line node at $k_z = 0$, whereas most experiments suggest that the nodes lie along the $\left[110\right]$ and $\left[\bar{1}10\right]$ directions \cite{lupien2002ultrasound,deguchi2004determination,HassingerPRX2017,sharma2019momentum}. There is some evidence, however, for a horizontal line node from angle-dependent heat capacity measurements \cite{KittakaJPSJ2018}. Second, \sro has very weak interlayer coupling \cite{BergemannPRL2000}, and in the limit of weak interlayer coupling, the pairing strength for this state goes to zero. A recent weak-coupling analysis shows that an $E_g$ state can be stabilized by including momentum-dependent spin orbit coupling \cite{AlinePRB2019,suh2019stabilizing}, and such spin-orbit coupling has been quantified by ARPES in \sro \cite{HaverkortPRL2008,VeenstraPRL2014,tamai2019high}. This variant of the $E_g$ state has Bogoliubov Fermi surfaces (rather than line nodes) that extend along the $k_z$ direction in a manner that may mimic line-nodes as far as experiment is concerned. 

The second possibility, $d_{x^2-y^2} \pm i g_{xy(x^2-y^2)}$, is less natural in that $d_{x^2-y^2}$ is a $B_{1g}$ irrep and $g_{xy(x^2-y^2)}$ is an $A_{2g}$ irrep \cite{kivelson2020proposal}. Order parameters of different representations have, in general, distinct transition temperatures, and therefore the composite B$_{1g}\oplus \mathrm{A}_{2g}$ order parameter requires fine-tuning to produce a single superconducting transition. Fine-tuning aside, this state has two attractive features. First, it produces bilinears only in the $A_{1g}$ and $B_{2g}$ channels ($B_{1g} \otimes  A_{2g} = B_{2g}$). This would naturally explain why a jump is seen in $c_{66}$ but not in $(c_{11} - c_{12})/2$. Second, this state has line nodes along the $\left[110\right]$ and $\left[\bar{1}10\right]$ directions \cite{lupien2002ultrasound,HassingerPRX2017,sharma2019momentum}. While the $l=4$, $g_{xy(x^2-y^2)}$ state may seem exotic, it has been shown (in the weak-coupling regime) to be competitive with $d_{x^2-y^2}$ when nearest-neighbor repulsion is accounted for \cite{Raghu:2012}. 

Both of these two-component order parameters produce a discontinuity in $c_{66}$, break time reversal symmetry, are Pauli limited in their upper critical field, exhibit a drop in the Knight shift below \Tc, and have ungapped quasiparticles. The accidental degeneracy of $d_{x^2-y^2} \pm i g_{xy(x^2-y^2)}$ means that its \Tc should split into two transitions under \textit{any} applied strain, and indeed, the aforementioned $\mu$SR measurements have found evidence for such a splitting \cite{grinenko2020split}. This suggests that \sro may indeed have two nearly degenerate transitions at ambient pressure. Whether or not such a fine-tuned state is tenable to all experiments remains to be seen, but it is now clear that a two-component superconducting order parameter is an essential feature for understanding the unusual superconducting properties of \sro.

\section{Methods}

\subsection*{Sample Preparation}
High-quality single crystal \sro were grown by the floating-zone method, details of which can be found in \cite{BobowskiCondMat2019}. A single crystal was oriented along the [110], [1$\bar{1}$0] and [001] directions, and polished to dimensions 1.50mm $\times$ 1.60mm $\times$ 1.44mm,  with 1.44mm along the tetragonal c-axis. The [110] orientation of the crystal was accounted for when solving for the elastic moduli.

\subsection*{Sample Characterization}

\begin{figure*}[h!]
	\centering
	\includegraphics[width=.99\linewidth]{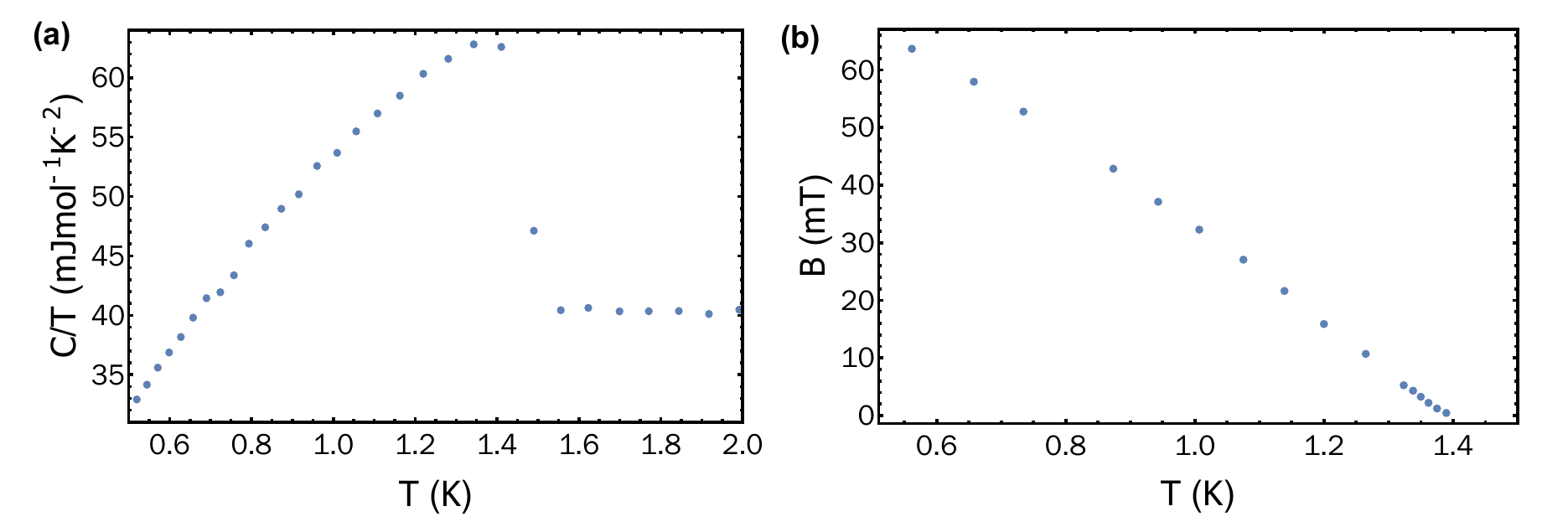}
	\caption{\textbf{Characterization of the \sro rod.} (a) Specific heat and (b) susceptibility measurements of the upper critical field, measured on different parts of the same rod from which the sample for RUS experiment was obtained. \Tc varies by about 200 mK between different parts of the rod.}
	\label{fig:sampleChar}
\end{figure*}
The quality of the \sro rod from which the RUS sample was cut was characterized by heat capacity and AC susceptibility measurements, shown in \autoref{fig:sampleChar}. Heat capacity measurements of a large piece of the rod exhibit a \Tc around 1.45 K, which is close to the optimal \Tc \cite{MackenziePRL1998}. In addition to a relatively high \Tc, a low concentration of ruthenium inclusions was an important criterion for the selection of the RUS sample. Ruthenium inclusions locally strain the crystal lattice and can enhance \Tc up to 3 K. In order to check for ruthenium inclusions, we measured AC susceptibility by a mutual inductance method, and found a sharp onset-Tc of 1.43 K, with no sign of a \Tc at 3 K, indicating a very low concentration of ruthenium inclusions. The variation in \Tc between the heat capacity and the susceptibility/critical field experiments arises because the samples were taken from different parts of the same \sro rod. The \sro sample for the RUS experiments was also taken from the same rod, and the onset \Tc of 1.425 K is in good agreement with the above-mentioned measurements.

\subsection*{Low-Temperature RUS}

The relatively low \Tc of \sro poses several challenges to perform RUS experiments through the superconducting transition. Since RUS samples are typically large (of the order 1 mm$^3$) and not glued to the transducers (to ensure free boundary conditions), there is weak thermal contact between the sample and its surroundings. Hence, when cooled through \Tc, the entire sample may not become superconducting at once, leading to broad superconducting transitions rather than sharp jumps at \Tc. To cool below 4.2 K, one could introduce liquid helium into the sample space and pump directly on this space. As RUS is extremely sensitive to vibration, however, this introduces artefacts into the data, and does not provide a particularly homogeneous thermal environment. 

\begin{figure*}[h!]
	\centering
	\includegraphics[width=.99\linewidth]{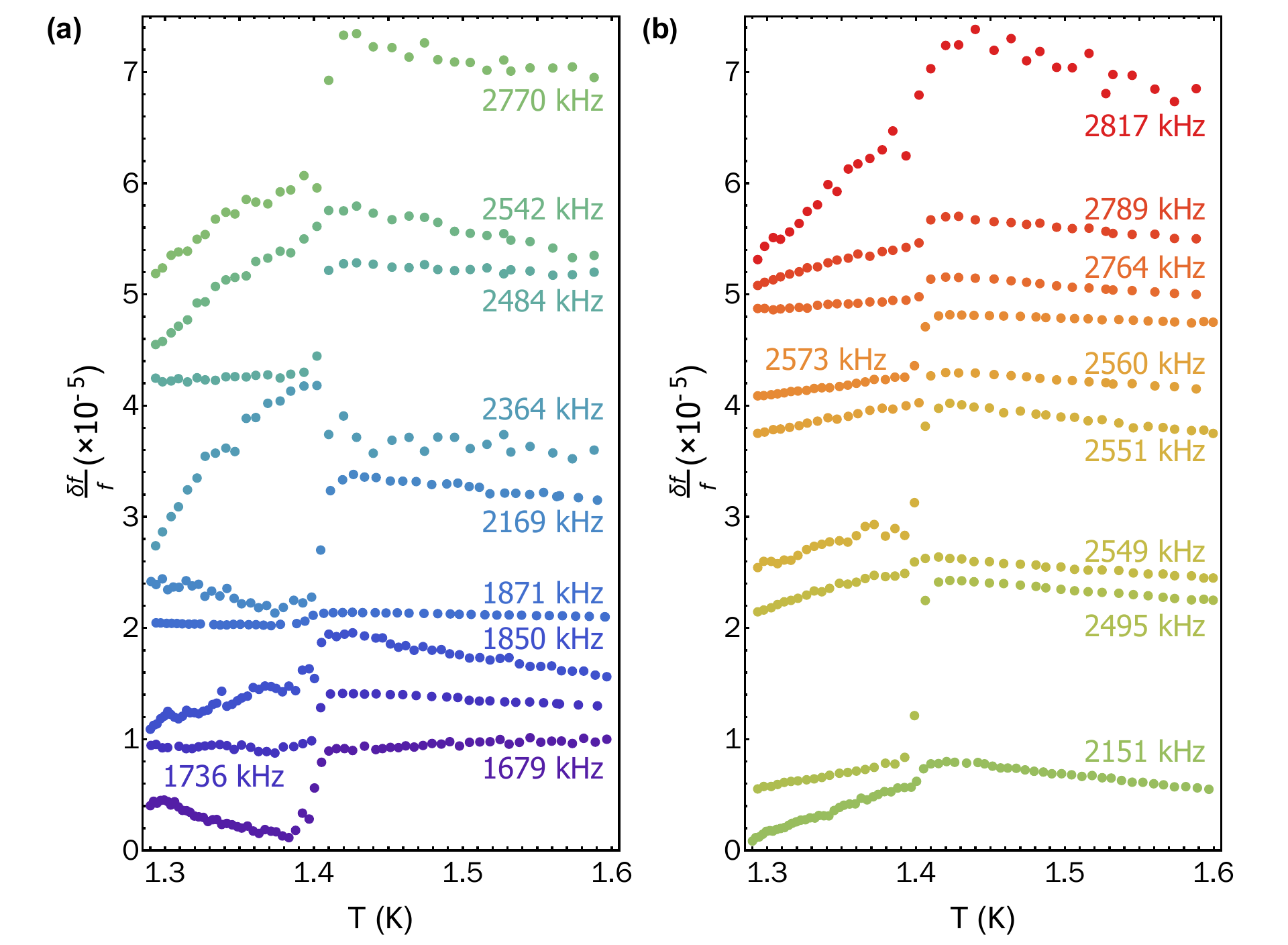}
	\caption{\textbf{RUS frequency data.} Temperature evolution of 18 resonance frequencies of \sro through \Tc, with panels (a) and (b) each showing 9 frequencies. Plots are vertically shifted for visual clarity.}
	\label{fig:fvT}
\end{figure*}

To solve these problems the RUS probe was sealed inside a copper can with a small amount of exchange gas, providing good thermal equilibration between the sample, thermometer, and the rest of the apparatus. This inner copper can is separated by a weak thermal link (thin-wall stainless) from an outer brass vacuum can, which provided isolation between the walls of the sample can and the bath. The temperature was regulated by pumping on the external helium bath, and the vacuum isolation of the sample chamber from the bath allows the sample space to then cool very slowly once the bath is pumped to base temperature. 

The lowest temperature reached was approximately 1.25 K, as read by a CX-1030 thermometer affixed to the RUS probe. The transition temperature and transition width observed by RUS  agree extremely well with those determined from independent susceptibility measurements, suggesting that the \sro sample was uniformly thermalized during the experiment.

We measured the temperature dependence of 18 resonances through the superconducting transition---the full data set is shown in \autoref{fig:fvT}. The temperature spacing is not identical because the data was acquired over several sweeps through \Tc, measuring a few resonances each time. These 18 resonances are then decomposed into the elastic moduli that are shown in Figure 3 of the main text. To do this decomposition we interpolate the frequency versus temperature data, and then plot the elastic moduli at what we consider to be a representative set of temperatures---those from the sweep where we measured the resonances at 2495 kHz, 2549 kHz, and 2551 kHz.

\section{Acknowledgments}

The authors acknowledge helpful discussions with Kimberly Modic, Steve Kivelson, Igor Mazin, Daniel Agterberg, Ronny Thomale, Peter Hirschfeld, Rafael Fernandes, Indranil Paul, Cyril Proust, and Louis Taillefer. 

B.J.R. and S.G. are grateful for help with the experimental design from Eric Smith, Jeevak Parpia, and the LASSP machine shop. B.J.R and S.G. acknowledge support for building the experiment, taking and analyzing the data, and writing the manuscript, from the U.S. Department of Energy, Office of Basic Energy Sciences under Award Number DE-SC0020143.  B.J.R. and S.G. acknowledge support by the Cornell Center for Materials Research with funding from the NSF MRSEC program (DMR-1719875). N.K. acknowledges the support from JSPS KAKENHI (No. JP18K04715) in Japan.

\section{Author Contributions}

S.G. and B.J.R designed the experiment. F.J., D.A.S., N.K., M.B., C.W.H. and A.P.M. prepared the crystal and performed characterization measurements. S.G. acquired and analyzed the ultrasound data. S.G., A.S., C.W.H. and B.J.R. wrote the manuscript with input from all co-authors.

\bibliographystyle{unsrtnat}
\bibliography{library}

\begin{thebibliography}{47}
\providecommand{\natexlab}[1]{#1}
\providecommand{\url}[1]{\texttt{#1}}
\expandafter\ifx\csname urlstyle\endcsname\relax
  \providecommand{\doi}[1]{doi: #1}\else
  \providecommand{\doi}{doi: \begingroup \urlstyle{rm}\Url}\fi

\bibitem[Rice and Sigrist(1995)]{RiceIOP1995}
T.~M. Rice and M.~Sigrist.
\newblock $\mathrm{Sr}_{2}\mathrm{RuO}_{4}$: an electronic analogue of
  $^3\mathrm{He}$?
\newblock \emph{Journal of Physics: Condensed Matter}, 7\penalty0
  (47):\penalty0 L643--L648, nov 1995.

\bibitem[Baskaran(1996)]{baskaran1996sr2ruo4}
G~Baskaran.
\newblock Why is sr2ruo4 not a high tc superconductor? electron correlation,
  hund's coupling and p-wave instability.
\newblock \emph{Physica B: Condensed Matter}, 223:\penalty0 490--495, 1996.

\bibitem[Maeno et~al.(1994)Maeno, Hashimoto, Yoshida, Nishizaki, Fujita,
  Bednorz, and Lichtenberg]{maeno1994superconductivity}
Y~Maeno, H~Hashimoto, K~Yoshida, S~Nishizaki, T~Fujita, JG~Bednorz, and
  F~Lichtenberg.
\newblock Superconductivity in a layered perovskite without copper.
\newblock \emph{Nature}, 372\penalty0 (6506):\penalty0 532--534, 1994.

\bibitem[Bergemann et~al.(2000)Bergemann, Julian, Mackenzie, Nishizaki, and
  Maeno]{BergemannPRL2000}
C.~Bergemann, S.~R. Julian, A.~P. Mackenzie, S.~Nishizaki, and Y.~Maeno.
\newblock Detailed topography of the fermi surface of
  ${\mathrm{sr}}_{2}{\mathrm{ruo}}_{4}$.
\newblock \emph{Phys. Rev. Lett.}, 84:\penalty0 2662--2665, Mar 2000.

\bibitem[Mackenzie and Maeno(2003)]{MackenzieRMP2003}
Andrew~Peter Mackenzie and Yoshiteru Maeno.
\newblock The superconductivity of $\mathrm{Sr}_{2}\mathrm{RuO}_{4}$ and the
  physics of spin-triplet pairing.
\newblock \emph{Rev. Mod. Phys.}, 75:\penalty0 657--712, May 2003.

\bibitem[Mackenzie et~al.(2017)Mackenzie, Scaffidi, Hicks, and
  Maeno]{MackenzieNPJ2017}
Andrew~P. Mackenzie, Thomas Scaffidi, Clifford~W. Hicks, and Yoshiteru Maeno.
\newblock Even odder after twenty-three years: the superconducting order
  parameter puzzle of {${\mathrm{Sr}}_{2}{\mathrm{RuO}}_{4}$}.
\newblock \emph{npj Quantum Materials}, 2\penalty0 (1):\penalty0 40, 2017.

\bibitem[Ishida et~al.(1998)Ishida, Mukuda, Kitaoka, Asayama, Mao, Mori, and
  Maeno]{IshidaNature1998}
K.~Ishida, H.~Mukuda, Y.~Kitaoka, K.~Asayama, Z.~Q. Mao, Y.~Mori, and Y.~Maeno.
\newblock Spin-triplet superconductivity in sr2ruo4 identified by 17o knight
  shift.
\newblock \emph{Nature}, 396\penalty0 (6712):\penalty0 658--660, 1998.

\bibitem[Pustogow et~al.(2019)Pustogow, Luo, Chronister, Su, Sokolov,
  Jerzembeck, Mackenzie, Hicks, Kikugawa, Raghu, Bauer, and
  Brown]{PustogowNat2019}
A.~Pustogow, Yongkang Luo, A.~Chronister, Y.-S. Su, D.~A. Sokolov,
  F.~Jerzembeck, A.~P. Mackenzie, C.~W. Hicks, N.~Kikugawa, S.~Raghu, E.~D.
  Bauer, and S.~E. Brown.
\newblock Constraints on the superconducting order parameter in sr2ruo4 from
  oxygen-17 nuclear magnetic resonance.
\newblock \emph{Nature}, 574\penalty0 (7776):\penalty0 72--75, 2019.

\bibitem[Ishida et~al.(2020)Ishida, Manago, Kinjo, and
  Maeno]{ishida2019reduction}
Kenji Ishida, Masahiro Manago, Katsuki Kinjo, and Yoshiteru Maeno.
\newblock Reduction of the 17o knight shift in the superconducting state and
  the heat-up effect by nmr pulses on sr2ruo4.
\newblock \emph{Journal of the Physical Society of Japan}, 89\penalty0
  (3):\penalty0 034712, 2020.

\bibitem[Kittaka et~al.(2009)Kittaka, Nakamura, Aono, Yonezawa, Ishida, and
  Maeno]{Kittaka:2009}
S.~Kittaka, T.~Nakamura, Y.~Aono, S.~Yonezawa, K.~Ishida, and Y.~Maeno.
\newblock Angular dependence of the upper critical field of
  ${\text{sr}}_{2}{\text{ruo}}_{4}$.
\newblock \emph{Phys. Rev. B}, 80:\penalty0 174514, Nov 2009.

\bibitem[Luke et~al.(1998)Luke, Fudamoto, Kojima, Larkin, Merrin, Nachumi,
  Uemura, Maeno, Mao, Mori, Nakamura, and Sigrist]{LukeNature1998}
G.~M. Luke, Y.~Fudamoto, K.~M. Kojima, M.~I. Larkin, J.~Merrin, B.~Nachumi,
  Y.~J. Uemura, Y.~Maeno, Z.~Q. Mao, Y.~Mori, H.~Nakamura, and M.~Sigrist.
\newblock Time-reversal symmetry-breaking superconductivity in sr2ruo4.
\newblock \emph{Nature}, 394\penalty0 (6693):\penalty0 558--561, 1998.

\bibitem[Xia et~al.(2006)Xia, Maeno, Beyersdorf, Fejer, and
  Kapitulnik]{XiaPRL2006}
Jing Xia, Yoshiteru Maeno, Peter~T. Beyersdorf, M.~M. Fejer, and Aharon
  Kapitulnik.
\newblock High resolution polar kerr effect measurements of
  ${\mathrm{sr}}_{2}{\mathrm{ruo}}_{4}$: Evidence for broken time-reversal
  symmetry in the superconducting state.
\newblock \emph{Phys. Rev. Lett.}, 97:\penalty0 167002, Oct 2006.

\bibitem[Jang et~al.(2011)Jang, Ferguson, Vakaryuk, Budakian, Chung, Goldbart,
  and Maeno]{JangScience2011}
J.~Jang, D.~G. Ferguson, V.~Vakaryuk, R.~Budakian, S.~B. Chung, P.~M. Goldbart,
  and Y.~Maeno.
\newblock Observation of half-height magnetization steps in sr2ruo4.
\newblock \emph{Science}, 331\penalty0 (6014):\penalty0 186--188, 2011.

\bibitem[Rehwald(1973)]{RehwaldAIP1973}
Walther Rehwald.
\newblock The study of structural phase transitions by means of ultrasonic
  experiments.
\newblock \emph{Advances in Physics}, 22\penalty0 (6):\penalty0 721--755, 1973.

\bibitem[Ghosh et~al.(2020)Ghosh, Matty, Baumbach, Bauer, Modic, Shekhter,
  Mydosh, Kim, and Ramshaw]{GhoshSciAdv2020}
Sayak Ghosh, Michael Matty, Ryan Baumbach, Eric~D. Bauer, K.~A. Modic, Arkady
  Shekhter, J.~A. Mydosh, Eun-Ah Kim, and B.~J. Ramshaw.
\newblock {One-component order parameter in
  ${\mathrm{U}\mathrm{R}\mathrm{u}_{2}\mathrm{S}\mathrm{i}_{2}}$ uncovered by
  resonant ultrasound spectroscopy and machine learning}.
\newblock \emph{Science Advances}, 6\penalty0 (10), 2020.

\bibitem[Okuda et~al.(2002)Okuda, Suzuki, Mao, Maeno, and
  Fujita]{okuda2002unconventional}
Noriki Okuda, Takashi Suzuki, Zhiqiang Mao, Yoshiteru Maeno, and Toshizo
  Fujita.
\newblock Unconventional strain dependence of superconductivity in spin-triplet
  superconductor sr2ruo4.
\newblock \emph{Journal of the Physical Society of Japan}, 71\penalty0
  (4):\penalty0 1134--1139, 2002.

\bibitem[Lupien(2002)]{lupien2002ultrasound}
Christian Lupien.
\newblock \emph{Ultrasound attenuation in the unconventional superconductor
  Sr2RuO4}.
\newblock PhD Thesis, 2002.

\bibitem[Walker and Contreras(2002)]{WalkerPRB2002}
M.~B. Walker and Pedro Contreras.
\newblock Theory of elastic properties of ${\mathrm{sr}}_{2}{\mathrm{ruo}}_{4}$
  at the superconducting transition temperature.
\newblock \emph{Phys. Rev. B}, 66:\penalty0 214508, Dec 2002.

\bibitem[Sigrist(2002)]{SigristPTP2002}
Manfred Sigrist.
\newblock {Ehrenfest Relations for Ultrasound Absorption in
  $\mathrm{Sr}_{2}\mathrm{RuO}_{4}$}.
\newblock \emph{Progress of Theoretical Physics}, 107\penalty0 (5):\penalty0
  917--925, 05 2002.

\bibitem[Migliori et~al.(1993)Migliori, Sarrao, Visscher, Bell, Lei, Fisk, and
  Leisure]{migliori1993resonant}
A.~Migliori, J.~L. Sarrao, William~M. Visscher, T.~M. Bell, Ming Lei, Z.~Fisk,
  and R~Gi Leisure.
\newblock Resonant ultrasound spectroscopic techniques for measurement of the
  elastic moduli of solids.
\newblock \emph{Physica B: Condensed Matter}, 183\penalty0 (1-2):\penalty0
  1--24, 1993.

\bibitem[Ramshaw et~al.(2015)Ramshaw, Shekhter, McDonald, Betts, Mitchell,
  Tobash, Mielke, Bauer, and Migliori]{RamshawPNAS}
B.~J. Ramshaw, Arkady Shekhter, Ross~D. McDonald, Jon~B. Betts, J.~N. Mitchell,
  P.~H. Tobash, C.~H. Mielke, E.~D. Bauer, and Albert Migliori.
\newblock Avoided valence transition in a plutonium superconductor.
\newblock \emph{Proceedings of the National Academy of Sciences}, 112\penalty0
  (11):\penalty0 3285--3289, 2015.

\bibitem[Forsythe et~al.(2002{\natexlab{a}})Forsythe, Julian, Bergemann, Pugh,
  Steiner, Alireza, McMullan, Nakamura, Haselwimmer, Walker, Saxena, Lonzarich,
  Mackenzie, Mao, and Maeno]{forsythe2002evolution}
D.~Forsythe, S.~R. Julian, C.~Bergemann, E.~Pugh, M.~J. Steiner, P.~L. Alireza,
  G.~J. McMullan, F.~Nakamura, R.~K.~W. Haselwimmer, I.~R. Walker, S.~S.
  Saxena, G.~G. Lonzarich, A.~P. Mackenzie, Z.~Q. Mao, and Y.~Maeno.
\newblock Evolution of fermi-liquid interactions in
  ${\mathrm{s}\mathrm{r}}_{2}{\mathrm{r}\mathrm{u}\mathrm{o}}_{4}$ under
  pressure.
\newblock \emph{Phys. Rev. Lett.}, 89:\penalty0 166402, Sep 2002{\natexlab{a}}.

\bibitem[{Grinenko} et~al.(2020){Grinenko}, {Ghosh}, {Sarkar}, {Orain},
  {Nikitin}, {Elender}, {Das}, {Guguchia}, {Br{\"u}ckner}, {Barber}, {Park},
  {Kikugawa}, {Sokolov}, {Bobowski}, {Miyoshi}, {Maeno}, {Mackenzie},
  {Luetkens}, {Hicks}, and {Klauss}]{grinenko2020split}
Vadim {Grinenko}, Shreenanda {Ghosh}, Rajib {Sarkar}, Jean-Christophe {Orain},
  Artem {Nikitin}, Matthias {Elender}, Debarchan {Das}, Zurab {Guguchia}, Felix
  {Br{\"u}ckner}, Mark~E. {Barber}, Joonbum {Park}, Naoki {Kikugawa}, Dmitry~A.
  {Sokolov}, Jake~S. {Bobowski}, Takuto {Miyoshi}, Yoshiteru {Maeno}, Andrew~P.
  {Mackenzie}, Hubertus {Luetkens}, Clifford~W. {Hicks}, and Hans-Henning
  {Klauss}.
\newblock {Split superconducting and time-reversal symmetry-breaking
  transitions, and magnetic order in Sr$_2$RuO$_4$ under uniaxial stress}.
\newblock \emph{arXiv e-prints}, art. arXiv:2001.08152, January 2020.

\bibitem[Hicks et~al.(2014)Hicks, Brodsky, Yelland, Gibbs, Bruin, Barber,
  Edkins, Nishimura, Yonezawa, Maeno, and Mackenzie]{HicksScience2014}
Clifford~W. Hicks, Daniel~O. Brodsky, Edward~A. Yelland, Alexandra~S. Gibbs,
  Jan A.~N. Bruin, Mark~E. Barber, Stephen~D. Edkins, Keigo Nishimura, Shingo
  Yonezawa, Yoshiteru Maeno, and Andrew~P. Mackenzie.
\newblock Strong increase of tc of sr2ruo4 under both tensile and compressive
  strain.
\newblock \emph{Science}, 344\penalty0 (6181):\penalty0 283--285, 2014.

\bibitem[{Li} et~al.(2019){Li}, {Kikugawa}, {Sokolov}, {Jerzembeck}, {Gibbs},
  {Maeno}, {Hicks}, {Nicklas}, and {Mackenzie}]{LiArxiv2019}
Y.~S. {Li}, N.~{Kikugawa}, D.~A. {Sokolov}, F.~{Jerzembeck}, A.~S. {Gibbs},
  Y.~{Maeno}, C.~W. {Hicks}, M.~{Nicklas}, and A.~P. {Mackenzie}.
\newblock {High precision heat capacity measurements on Sr2RuO4 under uniaxial
  pressure}.
\newblock \emph{arXiv e-prints}, art. arXiv:1906.07597, Jun 2019.

\bibitem[Watson et~al.(2018)Watson, Gibbs, Mackenzie, Hicks, and
  Moler]{watson2018micron}
Christopher~A. Watson, Alexandra~S. Gibbs, Andrew~P. Mackenzie, Clifford~W.
  Hicks, and Kathryn~A. Moler.
\newblock Micron-scale measurements of low anisotropic strain response of local
  ${T}_{c}$ in ${\mathrm{sr}}_{2}{\mathrm{ruo}}_{4}$.
\newblock \emph{Phys. Rev. B}, 98:\penalty0 094521, Sep 2018.

\bibitem[Deguchi et~al.(2004)Deguchi, Q.~Mao, and
  Maeno]{deguchi2004determination}
Kazuhiko Deguchi, Z~Q.~Mao, and Yoshiteru Maeno.
\newblock Determination of the superconducting gap structure in all bands of
  the spin-triplet superconductor sr $ \_2 $ ruo $ \_4$.
\newblock \emph{Journal of the Physical Society of Japan}, 73\penalty0
  (5):\penalty0 1313--1321, 2004.

\bibitem[Hassinger et~al.(2017)Hassinger, Bourgeois-Hope, Taniguchi, Ren\'e~de
  Cotret, Grissonnanche, Anwar, Maeno, Doiron-Leyraud, and
  Taillefer]{HassingerPRX2017}
E.~Hassinger, P.~Bourgeois-Hope, H.~Taniguchi, S.~Ren\'e~de Cotret,
  G.~Grissonnanche, M.~S. Anwar, Y.~Maeno, N.~Doiron-Leyraud, and Louis
  Taillefer.
\newblock Vertical line nodes in the superconducting gap structure of
  {${\mathrm{Sr}}_{2}{\mathrm{RuO}}_{4}$}.
\newblock \emph{Phys. Rev. X}, 7:\penalty0 011032, Mar 2017.

\bibitem[Sharma et~al.(2020)Sharma, Edkins, Wang, Kostin, Sow, Maeno,
  Mackenzie, Davis, and Madhavan]{sharma2019momentum}
Rahul Sharma, Stephen~D. Edkins, Zhenyu Wang, Andrey Kostin, Chanchal Sow,
  Yoshiteru Maeno, Andrew~P. Mackenzie, J.~C.~S{\'e}amus Davis, and Vidya
  Madhavan.
\newblock Momentum-resolved superconducting energy gaps of sr2ruo4 from
  quasiparticle interference imaging.
\newblock \emph{Proceedings of the National Academy of Sciences}, 117\penalty0
  (10):\penalty0 5222--5227, 2020.

\bibitem[R\o{}mer et~al.(2019)R\o{}mer, Scherer, Eremin, Hirschfeld, and
  Andersen]{romer2019knight}
A.~T. R\o{}mer, D.~D. Scherer, I.~M. Eremin, P.~J. Hirschfeld, and B.~M.
  Andersen.
\newblock Knight shift and leading superconducting instability from spin
  fluctuations in ${\mathrm{sr}}_{2}{\mathrm{ruo}}_{4}$.
\newblock \emph{Phys. Rev. Lett.}, 123:\penalty0 247001, Dec 2019.

\bibitem[Lupien et~al.(2001)Lupien, MacFarlane, Proust, Taillefer, Mao, and
  Maeno]{LupienPRL2001}
C.~Lupien, W.~A. MacFarlane, Cyril Proust, Louis Taillefer, Z.~Q. Mao, and
  Y.~Maeno.
\newblock Ultrasound attenuation in {${\mathrm{Sr}}_{2}{\mathrm{RuO}}_{4}$}: An
  angle-resolved study of the superconducting gap function.
\newblock \emph{Phys. Rev. Lett.}, 86:\penalty0 5986--5989, Jun 2001.

\bibitem[\ifmmode \check{Z}\else \v{Z}\fi{}uti\ifmmode~\acute{c}\else
  \'{c}\fi{} and Mazin(2005)]{zutic2005phase}
Igor \ifmmode \check{Z}\else \v{Z}\fi{}uti\ifmmode~\acute{c}\else \'{c}\fi{}
  and Igor Mazin.
\newblock Phase-sensitive tests of the pairing state symmetry in
  ${\mathrm{sr}}_{2}{\mathrm{ruo}}_{4}$.
\newblock \emph{Phys. Rev. Lett.}, 95:\penalty0 217004, Nov 2005.

\bibitem[Kittaka et~al.(2018)Kittaka, Nakamura, Sakakibara, Kikugawa,
  Terashima, Uji, Sokolov, Mackenzie, Irie, Tsutsumi, Suzuki, and
  Machida]{KittakaJPSJ2018}
Shunichiro Kittaka, Shota Nakamura, Toshiro Sakakibara, Naoki Kikugawa, Taichi
  Terashima, Shinya Uji, Dmitry~A. Sokolov, Andrew~P. Mackenzie, Koki Irie,
  Yasumasa Tsutsumi, Katsuhiro Suzuki, and Kazushige Machida.
\newblock Searching for gap zeros in {${\mathrm{Sr}}_{2}{\mathrm{RuO}}_{4}$}
  via field-angle-dependent specific-heat measurement.
\newblock \emph{Journal of the Physical Society of Japan}, 87\penalty0
  (9):\penalty0 093703, 2018.

\bibitem[Ramires and Sigrist(2019)]{AlinePRB2019}
Aline Ramires and Manfred Sigrist.
\newblock Superconducting order parameter of
  ${\mathrm{sr}}_{2}{\mathrm{ruo}}_{4}$: A microscopic perspective.
\newblock \emph{Phys. Rev. B}, 100:\penalty0 104501, Sep 2019.

\bibitem[Suh et~al.(2019)Suh, Menke, Brydon, Timm, Ramires, and
  Agterberg]{suh2019stabilizing}
Han~Gyeol Suh, Henri Menke, PMR Brydon, Carsten Timm, Aline Ramires, and
  Daniel~F Agterberg.
\newblock Stabilizing even-parity chiral superconductivity in sr $ \_2 $ ruo $
  \_4$.
\newblock \emph{arXiv preprint arXiv:1912.09525}, 2019.

\bibitem[Haverkort et~al.(2008)Haverkort, Elfimov, Tjeng, Sawatzky, and
  Damascelli]{HaverkortPRL2008}
M.~W. Haverkort, I.~S. Elfimov, L.~H. Tjeng, G.~A. Sawatzky, and A.~Damascelli.
\newblock Strong spin-orbit coupling effects on the fermi surface of
  ${\mathrm{sr}}_{2}{\mathrm{ruo}}_{4}$ and
  ${\mathrm{sr}}_{2}{\mathrm{rho}}_{4}$.
\newblock \emph{Phys. Rev. Lett.}, 101:\penalty0 026406, Jul 2008.

\bibitem[Veenstra et~al.(2014)Veenstra, Zhu, Raichle, Ludbrook, Nicolaou,
  Slomski, Landolt, Kittaka, Maeno, Dil, Elfimov, Haverkort, and
  Damascelli]{VeenstraPRL2014}
C.~N. Veenstra, Z.-H. Zhu, M.~Raichle, B.~M. Ludbrook, A.~Nicolaou, B.~Slomski,
  G.~Landolt, S.~Kittaka, Y.~Maeno, J.~H. Dil, I.~S. Elfimov, M.~W. Haverkort,
  and A.~Damascelli.
\newblock Spin-orbital entanglement and the breakdown of singlets and triplets
  in ${\mathrm{sr}}_{2}{\mathrm{ruo}}_{4}$ revealed by spin- and angle-resolved
  photoemission spectroscopy.
\newblock \emph{Phys. Rev. Lett.}, 112:\penalty0 127002, Mar 2014.

\bibitem[Tamai et~al.(2019)Tamai, Zingl, Rozbicki, Cappelli, Ricc\`o, de~la
  Torre, McKeown~Walker, Bruno, King, Meevasana, Shi,
  Radovi\ifmmode~\acute{c}\else \'{c}\fi{}, Plumb, Gibbs, Mackenzie, Berthod,
  Strand, Kim, Georges, and Baumberger]{tamai2019high}
A.~Tamai, M.~Zingl, E.~Rozbicki, E.~Cappelli, S.~Ricc\`o, A.~de~la Torre,
  S.~McKeown~Walker, F.~Y. Bruno, P.~D.~C. King, W.~Meevasana, M.~Shi,
  M.~Radovi\ifmmode~\acute{c}\else \'{c}\fi{}, N.~C. Plumb, A.~S. Gibbs, A.~P.
  Mackenzie, C.~Berthod, H.~U.~R. Strand, M.~Kim, A.~Georges, and
  F.~Baumberger.
\newblock High-resolution photoemission on
  ${\mathrm{sr}}_{2}{\mathrm{ruo}}_{4}$ reveals correlation-enhanced effective
  spin-orbit coupling and dominantly local self-energies.
\newblock \emph{Phys. Rev. X}, 9:\penalty0 021048, Jun 2019.

\bibitem[Kivelson et~al.(2020)Kivelson, Yuan, Ramshaw, and
  Thomale]{kivelson2020proposal}
Steven~Allan Kivelson, Andrew~Chang Yuan, Brad Ramshaw, and Ronny Thomale.
\newblock A proposal for reconciling diverse experiments on the superconducting
  state in sr2ruo4.
\newblock \emph{npj Quantum Materials}, 5\penalty0 (1):\penalty0 43, Jun 2020.

\bibitem[Raghu et~al.(2012)Raghu, Berg, Chubukov, and Kivelson]{Raghu:2012}
S.~Raghu, E.~Berg, A.~V. Chubukov, and S.~A. Kivelson.
\newblock Effects of longer-range interactions on unconventional
  superconductivity.
\newblock \emph{Phys. Rev. B}, 85:\penalty0 024516, Jan 2012.

\bibitem[Bobowski et~al.(2019)Bobowski, Kikugawa, Miyoshi, Suwa, Xu, Yonezawa,
  Sokolov, Mackenzie, and Maeno]{BobowskiCondMat2019}
Jake~S. Bobowski, Naoki Kikugawa, Takuto Miyoshi, Haruki Suwa, Han-shu Xu,
  Shingo Yonezawa, Dmitry~A. Sokolov, Andrew~P. Mackenzie, and Yoshiteru Maeno.
\newblock Improved single-crystal growth of sr2ruo4.
\newblock \emph{Condensed Matter}, 4\penalty0 (1), 2019.

\bibitem[Mackenzie et~al.(1998)Mackenzie, Haselwimmer, Tyler, Lonzarich, Mori,
  Nishizaki, and Maeno]{MackenziePRL1998}
A.~P. Mackenzie, R.~K.~W. Haselwimmer, A.~W. Tyler, G.~G. Lonzarich, Y.~Mori,
  S.~Nishizaki, and Y.~Maeno.
\newblock Extremely strong dependence of superconductivity on disorder in
  ${\mathrm{sr}}_{2}{\mathrm{ruo}}_{4}$.
\newblock \emph{Phys. Rev. Lett.}, 80:\penalty0 161--164, Jan 1998.

\bibitem[{Huang} et~al.(2019){Huang}, {Zhou}, and {Yao}]{HuangArxiv2019a}
Wen {Huang}, Yi~{Zhou}, and Hong {Yao}.
\newblock {Possible 3D nematic odd-parity pairing in Sr$_2$RuO$_4$:
  experimental evidences and predictions}.
\newblock \emph{arXiv e-prints}, art. arXiv:1901.07041, Jan 2019.

\bibitem[Shekhter et~al.(2013)Shekhter, Ramshaw, Liang, Hardy, Bonn, Balakirev,
  McDonald, Betts, Riggs, and Migliori]{ShekhterNat2013}
Arkady Shekhter, B.~J. Ramshaw, Ruixing Liang, W.~N. Hardy, D.~A. Bonn,
  Fedor~F. Balakirev, Ross~D. McDonald, Jon~B. Betts, Scott~C. Riggs, and
  Albert Migliori.
\newblock Bounding the pseudogap with a line of phase transitions in
  {YBa$_2\mathrm{Cu_3O_{6+\delta}}$}.
\newblock \emph{Nature}, 498:\penalty0 75 EP --, Jun 2013.

\bibitem[Nyhus et~al.(2002)Nyhus, Thisted, Kikugawa, Suzuki, and
  Fossheim]{Nyhus2002}
J.~Nyhus, U.~Thisted, N.~Kikugawa, T.~Suzuki, and K.~Fossheim.
\newblock Elastic and specific heat critical properties of la1.85sr0.15cuo4.
\newblock \emph{Physica C: Superconductivity}, 369\penalty0 (1):\penalty0 273
  -- 277, 2002.

\bibitem[Forsythe et~al.(2002{\natexlab{b}})Forsythe, Julian, Bergemann, Pugh,
  Steiner, Alireza, McMullan, Nakamura, Haselwimmer, Walker, Saxena, Lonzarich,
  Mackenzie, Mao, and Maeno]{ForsythePRL2002}
D.~Forsythe, S.~R. Julian, C.~Bergemann, E.~Pugh, M.~J. Steiner, P.~L. Alireza,
  G.~J. McMullan, F.~Nakamura, R.~K.~W. Haselwimmer, I.~R. Walker, S.~S.
  Saxena, G.~G. Lonzarich, A.~P. Mackenzie, Z.~Q. Mao, and Y.~Maeno.
\newblock Evolution of fermi-liquid interactions in
  ${\mathrm{s}\mathrm{r}}_{2}{\mathrm{r}\mathrm{u}\mathrm{o}}_{4}$ under
  pressure.
\newblock \emph{Phys. Rev. Lett.}, 89:\penalty0 166402, Sep 2002{\natexlab{b}}.

\bibitem[Fossheim and Berre(1972)]{FossheimPRB1972}
K.~Fossheim and B.~Berre.
\newblock Ultrasonic propagation, stress effects, and interaction parameters at
  the displacive transition in srti${\mathrm{o}}_{3}$.
\newblock \emph{Phys. Rev. B}, 5:\penalty0 3292--3308, Apr 1972.

\end{thebibliography}

\newpage
\section{Supplementary Information}
\subsection*{Thermal Homogeneity Below \Tc}

While the presence of exchange gas in the sample space ensures that all sample surfaces are at a uniform temperature (which is the same temperature read by the thermometer), we rely on good thermal conduction for the sample interior to equilibrate with the surfaces. Above \Tc, normal state \sro is a good metal which conducts heat well, and thus the entire volume of the sample should be in thermal equilibrium with surroundings. Below \Tc, however, temperature gradients may be enhanced due to the loss of heat carriers, as Cooper pairs do not carry heat. We expect this not to be an issue since \sro is a nodal superconductor, which means that there are always normal quasiparticles that carry heat, and the lowest temperature we reach is $\sim 0.8$\Tc, which means that the superconducting gap has only partially opened. 

To confirm that the thermal gradients in our sample are always small, we perform a simple calculation starting from the 3-dimensional heat flow equation,
\begin{equation}
\frac{\partial T}{\partial t}=\alpha\nabla^2T,
\label{eq:heateq}
\end{equation}
where $T(x,y,z,t)$ denotes the temperature profile within the sample, and $\alpha$ is the thermal diffusivity. For tetragonal \sro, this takes the form
\begin{equation}
\frac{\partial T}{\partial t}=\alpha_a\bigg(\frac{\partial^2T}{\partial x^2}+\frac{\partial^2T}{\partial y^2}\bigg)+\alpha_c\frac{\partial^2T}{\partial z^2}.
\label{eqn:heat}
\end{equation}
The thermal diffusivity $\alpha$ is related to the thermal conductivity $\kappa$ as $\alpha=\kappa/\rho C$, where $\rho$ is the density of \sro and $C$ is the specific heat. For \sro below \Tc, we have $C=87$ mJmol$^{-1}$K$^{-1}$, $\kappa_a=13.05$ Wm$^{-1}$K$^{-1}$ and $\kappa_c=0.07$ Wm$^{-1}$K$^{-1}$, giving $\alpha_a=8.61\times10^{-3}$ m$^{2}/$s and $\alpha_c=0.05\times10^{-3}$ m$^{2}/$s.

We first consider the effects of a step-change in the exchange-gas temperature as a worst-case scenario. With the boundary condition that all sample surfaces are at $T_0$, the solution of \autoref{eqn:heat} is
\begin{equation}
\begin{aligned}[b]
T(x,y,z,t)=T_0 + &\sum_{m,n,p=1}^{\infty}A_{mnp}\sin\bigg(\frac{m\pi x}{L_x}\bigg)\sin\bigg(\frac{n\pi y}{L_y}\bigg)\sin\bigg(\frac{p\pi z}{L_z}\bigg)\\&\boldsymbol{\cdot}\exp\bigg(-\pi^2\bigg(\frac{\alpha_am^2}{L_x^2}+\frac{\alpha_an^2}{L_y^2}+\frac{\alpha_cp^2}{L_z^2}\bigg)t\bigg),
\end{aligned}
\label{eqn:heatsol}
\end{equation}
where $m,n,p$ are integers, $L_x,L_y,L_z$ are the sample dimensions, and the coefficients $A_{mnp}$ depend on the initial temperature profile $T(x,y,z,t=0)$ within the sample. Thus, temperature variations from $T_0$ within the sample die out exponentially fast and, in the $t\rightarrow\infty$ limit, $T(x,y,z)\rightarrow T_0$. In particular, the slowest equilibration occurs when $A_{111}\neq0$ and all other $A_{mnp}=0$, since higher harmonic components have a faster exponential decay. In this case, \autoref{eqn:heatsol} simplifies to
\begin{equation}
T(x,y,z,t)=T_0+A_{111}\sin\bigg(\frac{\pi x}{L_x}\bigg)\sin\bigg(\frac{\pi y}{L_y}\bigg)\sin\bigg(\frac{\pi z}{L_z}\bigg)\boldsymbol{\cdot}\exp\bigg(-\pi^2\bigg(\frac{\alpha_a}{L_x^2}+\frac{\alpha_a}{L_y^2}+\frac{\alpha_c}{L_z^2}\bigg)t\bigg).
\end{equation}
The temperature in the middle of the sample ($T_{mid}(t)=T(L_x/2,L_y/2,L_z/2,t)$) differs most from $T_0$ at $t=0$, hence we can just look at $T_{mid}$ to get an upper bound on how long it takes for the entire sample to come to $T_0$.
\begin{equation}
T_{mid}(t)-T_0=\Delta T\exp\bigg(-\pi^2\bigg(\frac{\alpha_a}{L_x^2}+\frac{\alpha_a}{L_y^2}+\frac{\alpha_c}{L_z^2}\bigg)t\bigg)=\Delta T\exp(-\frac{t}{14~ \mu \mathrm{s} }),
\label{eqn:tmid}
\end{equation}
where we have used the dimensions of the sample (1.50mm $\times$ 1.60mm $\times$ 1.44mm), and we have used the worst-case diffusivity taken below \Tc from the thermal conductivity data reported in \citet{HassingerPRX2017} and using the specific heat measured on the same rod our sample came from (see \autoref{fig:sampleChar}). \autoref{eqn:tmid} shows that if a temperature difference $\Delta T$ appears within the sample, it reduces by a factor of 1000 in less than $10^{-4}$ s. For comparison, we cool the sample at $\approx0.3$ mK/s, and acquire a data point roughly ever 10 mK. Thus the equilibration time is much faster than time scale over which we do our measurements.

Another consideration is the total temperature offset in the center of the sample as we sweep the temperature, which is not captured by \autoref{eqn:tmid}. The steady-state solution of \autoref{eqn:heat} for a constant cooling rate results in parabolic temperature profile inside the sample. The steady state profile $T_{eq}$ is given by,
\begin{equation}
T_{eq}(x,y,z)=T_0+64\frac{x(L_x-x)y(L_y-y)z(L_z-z)}{L_x^2L_y^2L_z^2}\delta T
\end{equation}
Clearly, $T_{eq}=T_0$ at the sample surfaces, and the middle of sample is at $T_0+\delta T$. The offset $\delta T$ can be calculated by evaluating \autoref{eq:heateq} at $(x=L_x/2,y=L_y/2,z=L_z/2)$ and using the cooling rate ($\partial T/\partial t = 0.3$ mK/s),
\begin{equation}
\delta T=\frac{\partial T/\partial t}{8\Big(\frac{\alpha_a}{L_x^2}+\frac{\alpha_a}{L_y^2}+\frac{\alpha_c}{L_z^2}\Big)} \approx 6\hspace{1mm} \textrm{nK.}
\end{equation}

Immediately below \Tc, the heat capacity increases by approximately $50\%$. Over our full temperature range, the thermal conductivity $\kappa_a$ drops by about $10\%$~\cite{HassingerPRX2017}. The thermal diffusivity therefore drops by $\approx 50\%$ below \Tc. Even with this relatively large change in diffusivity, our results above demonstrate that our sample is very homogeneous in temperature during the course of the RUS experiment---both above and below \Tc.

\subsection*{Uncertainty Analysis }

While visual inspection is usually sufficient to determine whether or not a particular modulus shows a discontinuity at \Tc, numerical values are needed to perform consistency checks on the data, and to make quantitative predictions for future experiments. 

\begin{figure*}
	\centering
	\includegraphics[width=.95\linewidth]{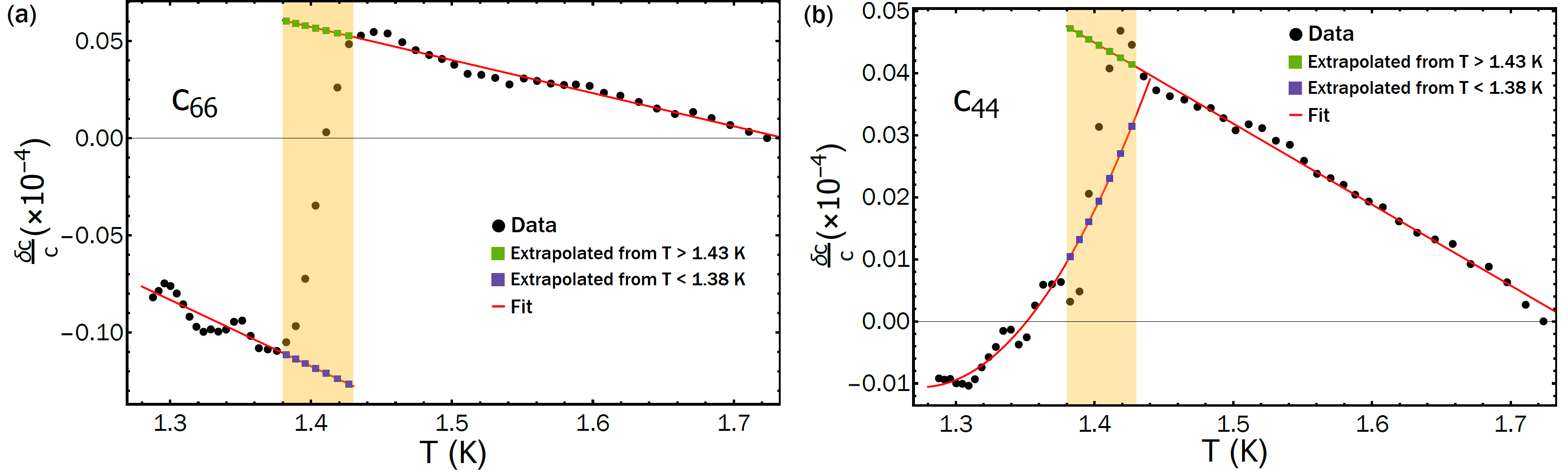}
	\caption{\textbf{Extraction of the jumps in the elastic moduli and their uncertainties.} We extrapolate fits to data (red lines) from above and below \Tc at temperature points within the transition (highlighted in yellow). The average between the minimum and maximum jump is taken to the be jump magnitude; the difference is the uncertainty. This procedure is illustrated for $c_{66}$ in panel (a), and $c_{44}$ in panel (b). Note the significantly reduced vertical scale in (b) as compared to (a).}
	\label{fig:jumpcalc}
\end{figure*}

One source of uncertainty comes from the width of the superconducting transition. To extract the jump magnitudes and their associated uncertainties in a consistent fashion for all the moduli, we use fits above and below \Tc to extrapolate the data across the transition (see \autoref{fig:jumpcalc}). We take the jump as the difference between the two extrapolated fits at the experimentally-obtained temperature points between 1.38 K and 1.43 K, which is the width of the superconducting transition. This gives a (non-Gaussian) distribution of modulus jumps that correspond to different \Tc assignments. We take the uncertainty to be half the difference of the minimum and maximum jumps, and assign the jump itself to the mean. 

A second source of uncertainty comes from the sample dimensions, which are used to extract the elastic moduli from the resonance frequencies. Slight deviations from parallelism, rounding of the sample corners, and other small imperfections give an upper-bound on the dimensional uncertainty of $\pm$20 microns in each direction. This dimensional uncertainty is then incorporated into the elastic moduli calculations (described in \citet{RamshawPNAS}), yielding an uncertainty in the moduli. This gives us the error bars shown for compressional and shear moduli in Figure 3(b) and (c) of main text. 

Assuming that the two sources of uncertainty (dimensional uncertainty, and the uncertainty in assigning \Tc) are independent, we add them in quadrature to obtain the total uncertainty in each jump, tabulated in \autoref{tab:errors}. 

We have neglected uncertainty due to misalignment of the crystal axes with respect to the sample faces because our crystal was aligned to better than $2^{\circ}$ for all 3 axes. The effect of misalignment can be calculated finding the rotated elastic tensor $\bm{C'}$ via
\begin{equation}
C'_{\text{mnop}} = R_{\text{mi}}R_{\text{nj}}R_{\text{ok}}R_{\text{pl}}C_{\text{ijkl}}
\label{eq:rot}
\end{equation}
where $\bm{C}$ is the un-rotated elastic tensor and $\bm{R}$ is a rotation matrix. For example, a rotation by an angle $\gamma$ about the $a$-axis transforms the shear elastic modulus $c_{44}$ as
\begin{equation}
c'_{44} = \frac{1}{4}\left[c_{44}\left(3+\cos4\gamma\right) + \frac{1}{2}\left(c_{11}+c_{33}+2c_{13}\left(1-\cos4\gamma\right)\right)\right].
\label{eq:rot2}
\end{equation}

For $\gamma = 2^{\circ}$, $c'_{44}$ differs from $c_{44}$ by four parts in $10^{3}$. This introduces a jump into $c'_{44}$ that is 1 part in $10^8$---two orders of magnitude smaller than the other sources of uncertainty. Similar expressions can be derived for the other moduli and are similarly small.

\begin{table*}
	\centering	
	\begin{tabular}{||c|| p{2cm}| p{1.1cm}| p{1.1cm}| p{2cm}| p{1.1cm}| p{1.1cm}|}
		\hline
		Elastic Modulus & 	$(c_{11}+c_{12})/2$ ($A_{1g}$) & $c_{33}$  ($A_{1g}$)  & $c_{13} $  ($A_{1g}$)   & $(c_{11}-c_{12})/2$  ($B_{1g}$)  &  $c_{44}$  ($E_{g}$)  & 	$c_{66}$  ($B_{2g}$) \\ 
		\hline
		Fractional Jump Size ($\times 10^{-5}$) & 5.35 & 3.79 & 10.74 & -0.07 & 0.22 & 1.75\\ 
		\hline	
		Uncertainty from \Tc width ($\times 10^{-5}$) & 0.09 & 0.02 & 0.10 & 0.49 & 0.13 & 0.04  \\
		\hline
		Uncertainty from Dimensions ($\times 10^{-5}$) & 0.24 & 0.39 & 0.72 & 0.41 & 0.44 & 0.24\\
		\hline
		\hline
		Total Uncertainty ($\times 10^{-5}$) & 0.25 & 0.39 & 0.72 & 0.64 & 0.46 & 0.25  \\
		\hline
	\end{tabular}
	\caption{Summary of jump magnitudes and experimental uncertainties in all six elastic moduli of \sro. The two sources of uncertainty---from width of superconducting transition and from sample dimensions---are added in quadrature to get the total uncertainty for each modulus. This leads to our error bars in Figure 3(d) of main text.}
	\label{tab:errors}
\end{table*}

\subsection*{Strain-Order Parameter Coupling}

In this section, we calculate expressions for the elastic moduli discontinuities and sound attenuation, starting from a Landau theory. Similar expressions for the chiral state have been calculated by \citet{SigristPTP2002} and for nematic superconducting states by \citet{HuangArxiv2019a}. Our expressions match those of Sigrist for the chiral state, and we correct the expressions derived by Huang et al. by including all possible order parameter fluctuations. In particular, we find that both in-plane shear moduli ($(c_{11}-c_{12})/2$ and $c_{66}$) should show a discontinuity for the nematic states, contrary to what was concluded in \cite{HuangArxiv2019a}. We also show that the three $A_{1g}$ jumps always follow a consistency relation, similar to what was concluded for non-superconducting order parameters in URu$_2$Si$_2$ \cite{GhoshSciAdv2020}, which is also a tetragonal material with point group $D_{4h}$.

For a two-component superconducting order parameter $\boldsymbol{\eta}=(\eta_x,\eta_y)$, the Landau free energy expansion reads
\begin{equation}
\mathcal{F}_{op}(\boldsymbol{\eta}) =a|\boldsymbol{\eta}|^2+b_1|\boldsymbol{\eta}|^4+\frac{b    _2}{2}\Big((\eta_x^*\eta_y)^2+(\eta_x\eta_y^*)^2\Big)+b_3|\eta_x|^2|\eta_y|^2 + ...
\label{eqn:Fop}
\end{equation}
where $a=a_0(T-T_c)$, with $a_0>0$, and $b_i$ are phenomenological constants.
In a tetragonal crystal, the elastic free energy density is given by
\begin{equation}
\begin{aligned}[b]
\mathcal{F}_{el} &=\frac{1}{2}\Big(c_{11}(\epsilon_{xx}^2+\epsilon_{yy}^2)+2c_{12}\epsilon_{xx}\epsilon_{yy}+c_{33}\epsilon_{zz}^2+2c_{13}(\epsilon_{xx}+\epsilon_{yy})\epsilon_{zz}+4c_{44}(\epsilon_{xz}^2+\epsilon_{yz}^2)+4c_{66}\epsilon_{xy}^2\Big) \\
&=\frac{1}{2}\Big(\frac{c_{11}+c_{12}}{2}(\epsilon_{xx}+\epsilon_{yy})^2+c_{33}\epsilon_{zz}^2+2c_{13}(\epsilon_{xx}+\epsilon_{yy})\epsilon_{zz}+\frac{c_{11}-c_{12}}{2}(\epsilon_{xx}-\epsilon_{yy})^2+\\&4c_{44}(\epsilon_{xz}^2+\epsilon_{yz}^2)+4c_{66}\epsilon_{xy}^2\Big) \\
&=\frac{1}{2}\Big(c_{A_{1g,1}}\epsilon_{A_{1g,1}}^2+c_{A_{1g,2}}\epsilon_{A_{1g,2}}^2+2c_{A_{1g,3}}\epsilon_{A_{1g,1}}\epsilon_{A_{1g,2}}+c_{B_{1g}}\epsilon_{B_{1g}}^2+c_{E_g}|\epsilon_{E_g}|^2+c_{B_{2g}}\epsilon_{B_{2g}}^2\Big)
\end{aligned}
\label{eqn:Felastic}
\end{equation}
where the strains are written as the irreducible representations of $D_{4h}$, $(\epsilon_{xx}+\epsilon_{yy}) \rightarrow \epsilon_{A_{1g,1}}$, $\epsilon_{zz} \rightarrow \epsilon_{A_{1g,2}}$, $(\epsilon_{xx}-\epsilon_{yy}) \rightarrow \epsilon_{B_{1g}}$, $2\epsilon_{xy} \rightarrow \epsilon_{B_{2g}}$ and $(2\epsilon_{xz},2\epsilon_{yz}) \rightarrow \epsilon_{E_g}$.

\begin{figure*}
	\centering
	\includegraphics[width=.99\linewidth]{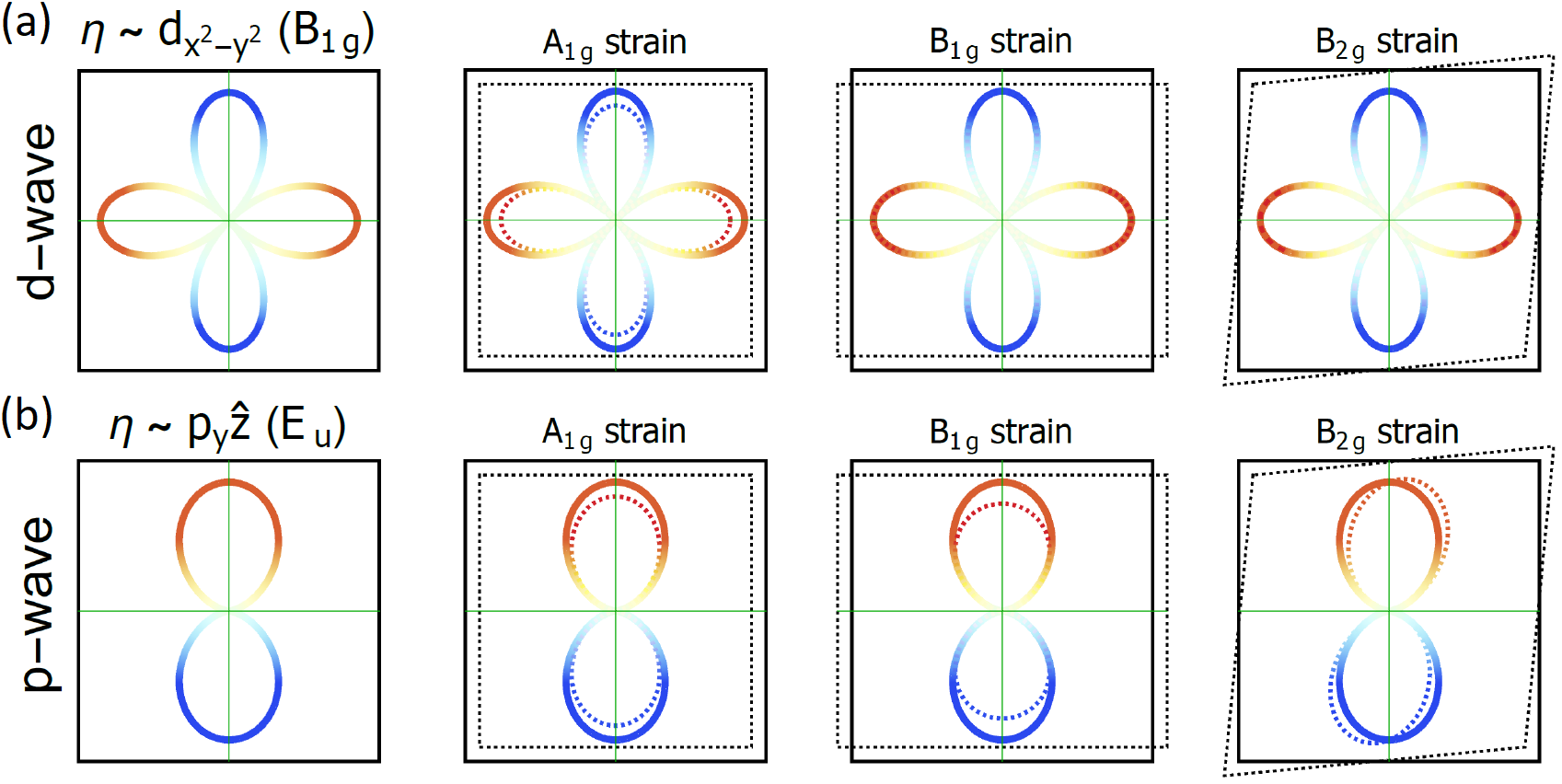}
	\caption{\textbf{Coupling between strain and one/two component order parameters.} (a) Structure of $d_{x^2-y^2}$ gap and (b) $p_y\hat{\boldsymbol{z}}$ gap in k-space, and their couplings to strain. $\hat{\boldsymbol{z}}$ represents the pair wavefunction in spin space. Allowed couplings modify the gap structure in k-space. Only $A_{1g}$ strain couples to one-component order parameters, while $A_{1g}$, $B_{1g}$ and $B_{2g}$ strain all couple to two-component order parameters.}
	\label{fig:couplingSI}
\end{figure*}

The lowest order terms that couple strain to the superconducting order parameter must be quadratic in order parameter to preserve gauge symmetry. This coupling gives rise to additional contributions to the free energy
\begin{equation}
\mathcal{F}_{c} =(g_1\epsilon_{A_{1g,1}}+g_2\epsilon_{A_{1g,2}})|\boldsymbol{\eta}|^2+g_4\epsilon_{B_{1g}}(|\eta_x|^2-|\eta_y|^2)+g_5\epsilon_{B_{2g}}(\eta_x^*\eta_y+\eta_x\eta_y^*),
\label{eqn:Fc}
\end{equation}
where $g_i$ are coupling constants. Coupling between OP and $B_{1g}$, $B_{2g}$ strains are only allowed for two-component OPs; one-component OPs can only couple to compressive strains (shown for example gap structures in \autoref{fig:couplingSI}). These linear-in-strain, quadratic-in-order-parameter coupling lead to elastic moduli jumps at \Tc. Hence jumps in shear moduli can only occur if the OP is two-component. Since no OP can couple to $E_g$ strain, $c_{44}$ should not show a jump at \Tc for any superconducting order parameter.

Following Sigrist\cite{SigristPTP2002}, we use the parameterization $\boldsymbol{\eta}=(\eta_x,\eta_y)=\eta(\cos{\theta},e^{i\gamma}\sin{\theta})$. Depending on the relative magnitudes of $b_1,b_2,b_3$, the system can have different equilibrium OPs\cite{HuangArxiv2019a}, characterized by different equilibrium values of $(\theta,\gamma)=(\theta_0,\gamma_0)$: $(\pi/4,\pm\pi/2)$ for the chiral state, $(\pi/4,0)$ for the diagonal nematic state and $(0,0)$ for the horizontal nematic state. These states also have different equilibrium values of $\eta=\eta_0$, which can be calculated from $\partial\mathcal{F}_{op}/\partial\eta|_{(\theta,\gamma)\rightarrow(\theta_0,\gamma_0)}=0$. Fluctuations of the order parameter amplitude $\eta$, orientation $\theta$ or relative phase $\gamma$ can couple to different strains, leading to the jump in corresponding moduli.

To explicitly calculate the moduli discontinuities, we adapt the approach described in \cite{ShekhterNat2013} to a multi-component OP. For one-component OPs, the discontinuity is
\begin{equation}
c^{<}_{mn}=c^{>}_{mn}-\frac{Z_mZ_n}{Y}\implies \Delta c_{mn}=c^{>}_{mn}-c^{<}_{mn}=\frac{Z_mZ_n}{Y}
\label{eqn:jump}
\end{equation}
where $c^{<}_{mn}$($c^{>}_{mn}$) is the elastic modulus below(above) \Tc. The thermodynamic coefficients $Z_i$, $Y$ are defined as $Z_i=\partial^2\mathcal{F}_c/\partial{\eta}\partial{\epsilon_i}$ and $Y=\partial^2\mathcal{F}_{op}/\partial\eta^2$.
For a multi-component OP $\boldsymbol{\eta}$, \autoref{eqn:jump} gets modified to
\begin{equation}
\Delta{c_{mn}}=\boldsymbol{Z}_{\boldsymbol{m}}^T\boldsymbol{Y}^{-1}\boldsymbol{Z_n}
\label{eqn:modjump}
\end{equation}
where $\boldsymbol{Z_i}=\partial^2\mathcal{F}_c/\partial{\boldsymbol{\eta}}\partial{\epsilon_i}$ and $\boldsymbol{Y}=\partial^2\mathcal{F}_{op}/\partial{\boldsymbol{\eta}^2}$ are now matrices. Within this formalism, one can find which OP fluctuation mode couples to a particular strain by looking at the $Z_i$ for that strain. For example, for the chiral state, 

\begin{equation}
Z_{A_{1g},1(2)}=\begin{pmatrix}
g_{1(2)}\sqrt{\frac{-8a}{4b_1-b_2+b_3}}\\ 0\\ 0\\
\end{pmatrix};
Z_{B_{1g}}=\begin{pmatrix}
0\\ g_4\frac{4a}{4b_1-b_2+b_3}\\ 0\\
\end{pmatrix};
Z_{B_{2g}}=\begin{pmatrix}
0\\ 0\\ g_5\frac{2a}{4b_1-b_2+b_3}\\
\end{pmatrix}
\end{equation} 
This shows that $\eta$ fluctuations couple to the $A_{1g}$ strains, $\theta$ fluctuations couple to $B_{1g}$ strain and $\gamma$ fluctuations couple to $B_{2g}$ strain, consistent with the conclusions of \citet{SigristPTP2002}.

The elastic moduli discontinuities for the various OPs is summarized in \autoref{tab:summary}. In all cases, the three $A_{1g}$ jumps are found to satisfy the relation
\begin{equation}
\Delta{c_{A_{1g},1}} \times \Delta{c_{A_{1g},2}} =(\Delta{c_{A_{1g},3}})^2
\label{eqn:a1gjumps}
\end{equation} 
Using the measured jumps in $(c_{11}+c_{12})/2$, $c_{33}$, and $c_{13}$ (see Figure 3 of the main text and \autoref{eqn:Felastic}), we obtain $(\Delta \frac{c_{11} +c_{12}}{2}) \times (\Delta c_{33}) = (9.9\pm1.5)\!\times\! 10^{-5}~ \mathrm{GPa}^2$, which is in agreement with $(\Delta c_{13})^2 = (8.3\pm1.1)\!\times\! 10^{-5}~ \mathrm{GPa}^2$.

\begin{table*}
	\centering	
	\begin{tabular}{||c||c|c|c||}
		\hline
		OP & Chiral & Diagonal Nematic & Horizontal Nematic \\ 
		\hline\hline
		$(\theta_0,\gamma_0)$ & $(\pi/4,\pm\pi/2)$ & $(\pi/4,0)$ & $(0,0)$  \\ 
		\hline	
		$\eta_0$ & $\sqrt{\frac{-2a}{4b_1-b_2+b_3}}$ & $\sqrt{\frac{-2a}{4b_1+b_2+b_3}}$ & $\sqrt{\frac{-a}{2b_1}}$  \\
		\hline
		$\Delta{c_{A_{1g},1}}$ & $\frac{2g_1^2}{4b_1-b_2+b_3}$ $(\eta)$& $\frac{2g_1^2}{4b_1+b_2+b_3}$ $(\eta)$ & $\frac{g_1^2}{2b_1}$ $(\eta)$ \\
		\hline
		$\Delta{c_{A_{1g},2}}$ & $\frac{2g_2^2}{4b_1-b_2+b_3}$ $(\eta)$& $\frac{2g_2^2}{4b_1+b_2+b_3}$ $(\eta)$ & $\frac{g_2^2}{2b_1}$ $(\eta)$  \\
		\hline
		$\Delta{c_{A_{1g},3}}$ & $\frac{2g_1g_2}{4b_1-b_2+b_3}$ $(\eta)$& $\frac{2g_1g_2}{4b_1+b_2+b_3}$ $(\eta)$& $\frac{g_1g_2}{2b_1}$ $(\eta)$ \\
		\hline
		$\Delta{c_{B_{1g}}}$ & $\frac{2g_4^2}{b_2-b_3}$ $(\theta)$& $\frac{-2g_4^2}{b_2+b_3}$ $(\theta)$ & $\frac{g_4^2}{2b_1}$ $(\eta)$ \\
		\hline
		$\Delta{c_{B_{2g}}}$ & $\frac{g_5^2}{b_2}$ $(\gamma)$& $\frac{2g_5^2}{4b_1+b_2+b_3}$ $(\eta)$ & $\frac{2g_5^2}{b_2+b_3}$ $(\theta)$ \\
		\hline		
	\end{tabular}
	\caption{Different equilibrium order parameters and the discontinuities they produce in various elastic moduli. In parentheses, we note fluctuation of which OP mode couples to ultrasound. Jump in compressional ($A_{1g}$) moduli is always caused by amplitude ($\eta$) fluctuations of the OP, whereas for shear modes, jumps can arise from coupling to amplitude ($\eta$), orientation ($\theta$) or relative phase ($\gamma$) fluctuations.}
	\label{tab:summary}
\end{table*}

We now turn to order parameter dynamics near the phase transition, which can cause smearing of the frequency jumps. We start with the idea outlined in \cite{ShekhterNat2013}, and adapt it to a multi-component OP. Below \Tc, the order parameter relaxation can be modeled as 
\begin{equation}
\frac{\partial{\boldsymbol{\tilde{\eta}}}}{\partial{t}}=-\boldsymbol{\xi}\frac{\partial{\mathcal{F}}}{\partial{\boldsymbol{\tilde{\eta}}}}=-\boldsymbol{\xi}\Big(\frac{\partial^2\mathcal{F}}{\partial{\boldsymbol{\eta}}^2}\boldsymbol{\tilde{\eta}} +\sum_{m} \frac{\partial^2\mathcal{F}}{\partial{\boldsymbol{\eta}}\partial{\epsilon_m}}\epsilon_m \Big)=-\boldsymbol{\xi}\Big(\boldsymbol{Y}\boldsymbol{\tilde{\eta}} +\sum_{m} \boldsymbol{Z_m }\epsilon_m \Big)
\label{eqn:dyn}
\end{equation}
where $\boldsymbol{\tilde{\eta}}$ are the fluctuations of OP components about equilibrium, and $-\boldsymbol{\xi}\frac{\partial{\mathcal{F}}}{\partial{\boldsymbol{\tilde{\eta}}}}$ provides the restoring force towards equilibrium. Assuming linear response of OP fluctuations to strain, when strain is modulated at frequency $\omega$, as in a RUS experiment, \autoref{eqn:dyn} becomes
\begin{equation}
-i\omega\boldsymbol{\tilde{\eta}}(\omega)=-\boldsymbol{\tau}^{-1}\boldsymbol{\tilde{\eta}}(\omega)-\boldsymbol{\xi}\sum_{m}\boldsymbol{Z_m }\epsilon_m \implies \boldsymbol{\tilde{\eta}}(\omega)=(i\omega\boldsymbol{\tau}-\mathbb{1})^{-1}\boldsymbol{Y}^{-1}\sum_{m}\boldsymbol{Z_m }\epsilon_m
\label{eqn:dyn2}
\end{equation}
where we have defined $\boldsymbol{\tau}^{-1}=\boldsymbol{\xi}\boldsymbol{Y}$ is the matrix of relaxation times for independent OP modes. For the parameterization $\boldsymbol{\eta}=\eta(\cos{\theta},e^{i\gamma}\sin{\theta})$, 
\begin{equation}
\boldsymbol{\tau}=\begin{pmatrix}
\tau_{\eta} & 0 & 0\\ 0 & \tau_{\theta} & 0\\ 0 & 0 & \tau_{\gamma}\\
\end{pmatrix} \implies (i\omega\boldsymbol{\tau}-\mathbb{1})^{-1}=\begin{pmatrix}
(i\omega{\tau_{\eta}}-1)^{-1} & 0 & 0\\ 0 & (i\omega{\tau_{\theta}}-1)^{-1} & 0\\ 0 & 0 & (i\omega{\tau_{\gamma}}-1)^{-1}\\
\end{pmatrix}
\end{equation}
Using \autoref{eqn:dyn2}, we calculate the dynamic elastic constant as,
\begin{equation}
c_{mn}(\omega)=c_{mn}^{>}+\frac{\partial^2\mathcal{F}}{\partial{\boldsymbol{\eta}}\partial{\epsilon_m}}\frac{\partial{\boldsymbol{\tilde{\eta}}}}{\partial{\epsilon_n}}\implies c_{mn}^{<}(\omega)=c_{mn}^{>}+\boldsymbol{Z}_{\boldsymbol{m}}^T(i\omega\boldsymbol{\tau}-\mathbb{1})^{-1}\boldsymbol{Y}^{-1}\boldsymbol{Z_n}
\label{eqn:dyn3}
\end{equation}
Elastic moduli jumps come from the real part of \autoref{eqn:dyn3}. Depending on which OP mode a particular elastic moduli couples to, the modulus dispersion $c_{mn}(\omega)$ picks up a contribution from the corresponding relaxation time. For example, for the chiral OP, 
$$\Delta c_{A_{1g},1(2)}=\frac{2g_{1(2)}^2}{4b_1-b_2+b_3}\frac{1}{1+\omega^2\tau_{\eta}^2};\Delta c_{B_{1g}}=\frac{2g_4^2}{b_2-b_3}\frac{1}{1+\omega^2\tau_{\theta}^2};\Delta c_{B_{2g}}=\frac{g_5^2}{b_2}\frac{1}{1+\omega^2\tau_{\gamma}^2}$$  

Thus OP relaxation effects can broaden out the elastic moduli jumps, if particular relaxation times are long compared to the experimental frequencies. This has been observed experimentally, for example, in the cuprate superconductor LSCO\cite{Nyhus2002}---higher frequencies reduce the magnitude of jump measured. Since we measure non-zero discontinuities in all the $A_{1g}$ moduli, and the $B_{2g}$ modulus, but no jump in $B_{1g}$ modulus, it is plausible that the $B_{1g}$ jump gets strongly smeared due to this effect. Specifically, for the diagonal nematic state, only the $B_{1g}$ jump is affected by $\tau_{\theta}$, while the other 4 jumps are related to $\tau_{\eta}$. Hence if $\tau_{\theta}$ is much larger than $\tau_{\eta}$ for this state, it provides an explanation for the lack of $B_{1g}$ jump. We also note that for this state, close to \Tc ($1-T/T_c\ll 1$), $\tau_{\eta}^{-1}\propto|a|\propto \eta_0^2 \propto (T_c-T)$ and $\tau_{\theta}^{-1}\propto\eta_0^4\propto (T_c-T)^{2}$. This would indeed make $\tau_{\theta}$ much longer than $\tau_{\eta}$ just below \Tc. We also note that large ultrasonic attenuation is observed experimentally in the $B_{1g}$ channel \cite{lupien2002ultrasound}, which perhaps motivates the presence of such a long relaxation time.

\subsection*{Reconciling Resonant Ultrasound and Pulse Echo Ultrasound Experiments}

The $c_{66}$ discontinuity we measure with RUS is about a factor of 50 larger than what was measured with pulse-echo experiments\cite{lupien2002ultrasound}. This apparent discrepancy may be resolved by looking at the frequency scales of the two experiments: $\sim 2.5$ MHz for RUS versus $169$ MHz for pulse-echo. As noted in the previous section, higher frequencies are expected to reduce the magnitude of the experimentally measured jump. This has been observed at the superconducting transition in La$_{1.85}$Sr$_{0.15}$CuO$_4$, where the jump in $c_{33}$ decreases by a factor of $\sim 4$ when the measurement frequency is increased from 16 MHz to 214 MHz. 

Applying a simple relaxation model, like the one derived in the previous section for order-parameter relaxation near the phase transition (\autoref{eqn:dyn3}), we obtain 
\begin{equation}
\frac{\Delta c(2.5\: \mathrm{MHz})}{\Delta c(169 \: \mathrm{MHz})}=\frac{(1+(2.5 \: \mathrm{MHz}\cdot\tau)^2)^{-1}}{(1+(169 \: \mathrm{MHz}\cdot\tau)^2)^{-1}}=50 \implies \tau \sim 6.6 \: \mathrm{ns}
\end{equation} 
where $\tau$ is a relaxation timescale. Applying the same model to the La$_{1.85}$Sr$_{0.15}$CuO$_4$ data from \citet{Nyhus2002} we obtain $1$ ns --- a comparable timescale. Whatever the microscopic mechanism underlying this timescale, it is clear that lower-frequency measurements should measure a jump that is closer to the intrinsic thermodynamic jump.

\subsection*{Ehrenfest Relations for Compressional Strains}

At the superconducting transition, a jump discontinuity is also measured in the specific heat. Within our formalism, the specific heat jump at \Tc, $\Delta C/T$, is calculated as 
\begin{equation}
\frac{\Delta C}{T} =\boldsymbol{W}^T\boldsymbol{Y}^{-1}\boldsymbol{W}
\label{eqn:Cvjump} 
\end{equation}
where $\boldsymbol{W}=\partial^2\mathcal{F}_{op}/\partial{\boldsymbol{\eta}}\partial{T}$ and $\boldsymbol{Y}=\partial^2\mathcal{F}_{op}/\partial{\boldsymbol{\eta}^2}$. For all three superconducting states discussed above, the $A_{1g}$ moduli jumps can be related to the specific heat jump through
\begin{equation}
\Delta c_{A_{1g},1(2)}=-\frac{\Delta C}{T}\bigg(\frac{dT_c}{d\epsilon_{A_{1g,1(2)}}}\bigg)^2;\Delta c_{A_{1g},3}=-\frac{\Delta C}{T}\bigg|\frac{dT_c}{d\epsilon_{A_{1g,1}}}\bigg|\bigg|\frac{dT_c}{d\epsilon_{A_{1g,2}}}\bigg|
\label{eqn:ehren2} 
\end{equation}
It is important to note that such relations for the shear strains are more complicated, and we derive them in the next section. For a tetragonal material, the bulk modulus $B$ is related to the three $A_{1g}$ moduli as
\begin{equation}
B=\frac{\big(\frac{c_{11}+c_{12}}{2}\big)c_{33}-c_{13}^2}{\big(\frac{c_{11}+c_{12}}{2}\big)+c_{33}-2c_{13}}
\end{equation}
From \autoref{eqn:ehren2},the discontinuity in the bulk modulus $\Delta B/B$ at \Tc can be related to $\Delta C/T$ through the Ehrenfest relation
\begin{equation}
\frac{\Delta B}{B^2}=-\frac{\Delta C}{T}\bigg(\frac{dT_c}{dP_{hyd}}\bigg)^2
\end{equation}
where $dT_{c}/dP_{hyd}$ is the hydrostatic pressure dependence of \Tc. 
Our measurements give $\Delta B/B\sim 6.3\times10^{-5}$, about 9 times larger than estimated from specific heat jump and $dT_c/dP_{hyd}$ \cite{ForsythePRL2002} or, alternatively, the measured jump in the bulk modulus predicts $dT_c/dP_{hyd}$ to be a factor of 3 higher than the measured value. This discrepancy can arise due to the formation of order parameter domains \cite{FossheimPRB1972}, which lead to an additional slowing down of ultrasound and therefore a larger drop in the elastic moduli through \Tc. Since these domains would be related to each other by time-reversal in a superconductor, however, it is unclear whether such a mechanism would couple strongly to ultrasound. Another possible cause could be the value of $dT_c/dP_{hyd}$ estimated from the data in \cite{ForsythePRL2002}. Since the \Tc of \sro shows a strong increase with $B_{1g}$ strain\cite{HicksScience2014}, the measured decrease in \Tc under $P_{hyd}$ will be less if the pressure applying medium is not completely hydrostatic. This is particularly relevant because the $B_{1g}$ modulus is almost 4 times smaller than $(c_{11}+c_{12})/2$, which makes it easy to induce $B_{1g}$ strain if the pressure medium is not hydrostatic. Finally, with the possible discovery of two transitions occurring either simultaneously or near-simultaneously at \Tc, it will be necessary to map out $T$ \mbox{\tiny $\!\!\!\!_{TRSB}$ } with pressure to correctly calculate the Ehrenfest relations, which are modified under the presence of two accidentally degenerate order parameters \cite{kivelson2020proposal}.

\subsection*{Ehrenfest Relations for Shear Strains}

Unlike $A_{1g}$ strains, shear strains ($B_{1g}$ and $B_{2g}$) are expected to split the superconducting transition if the OP is a symmetry-protected multi-component order parameter \cite{HicksScience2014}. This happens because shear strains break the tetragonal symmetry of the lattice, and hence the degenerate OP components of the unstrained crystal now have different condensation energies (and temperatures). Within weak coupling, a crystal under shear strain should therefore show two specific heat jumps\cite{LiArxiv2019}, and, for a chiral OP, time-reversal symmetry breaking (TRSB) should set in at a different temperature than Meissner effect. Recent $\mu$SR experiments\cite{grinenko2020split} have indeed reported the latter effect. We show that the shear modulus jump can be related to $\Delta C/T$ (at zero strain) through the strain derivatives of these two transition temperatures, \Tc and $T$ \mbox{\tiny $\!\!\!\!_{TRSB}$},
\begin{equation}
\Delta c_{s} = -\frac{\Delta C}{T} \bigg|\frac{dT_1}{d\epsilon_{s}}\bigg|\bigg|\frac{dT_2}{d\epsilon_{s}}\bigg|,
\label{eqn:ehrensh}
\end{equation}
where $s$ is either $B_{1g}$ or $B_{2g}$, $T_1=T_c$, and $T_2=T$ \mbox{\tiny $\!\!\!\!_{TRSB}$}.

We start from the free energy expressions $\mathcal{F}_{op}$ and $\mathcal{F}_{c}$, and consider the case $b_2>0, b_3<b_2$, which leads to a chiral OP. Further, we keep only the coupling to $\epsilon_{B_{1g}}$ to simplify the subsequent algebra. Then, with the phase between $\eta_x$ and $\eta_y$ set to $\pi/2$, $\mathcal{F}_{op}$ and $\mathcal{F}_c$ are
\begin{equation}
\begin{aligned}[b]
\mathcal{F}_{op} &=a_0(T-T_{c,0})(\eta_x^2+\eta_y^2)+b_1(\eta_x^2+\eta_y^2)^2+(b_3-b_2)\eta_x^2\eta_y^2\\
\mathcal{F}_c&=g_4\epsilon_{B_{1g}}(\eta_x^2-\eta_y^2),
\end{aligned}
\label{eqn:Fnew}
\end{equation}
where $T_{c,0}$ is the unstrained \Tc. Clearly, $\epsilon_{B_{1g}}$ breaks the $\eta_x\leftrightarrow\eta_y$ symmetry of the quadratic terms in free energy, thereby making the two components condense at different temperatures. 

We assume $g_4\epsilon_{B_{1g}}>0$, which favors $\eta_y$ condensing before $\eta_x$. The higher transition temperature $T_1=$ \Tc is determined by when the coefficient of $\eta_y^2$ goes to zero (with $\eta_x=0$), that is, $a_0(T_1-T_{c,0})-g_4\epsilon_{B_{1g}}=0$. This gives
\begin{equation}
T_1=T_{c,0}+\frac{g_4}{a_0}\epsilon_{B_{1g}}.
\label{eqn:Tcplus}
\end{equation}
Then, the $\eta_y$ that minimizes $(\mathcal{F}_{op}+\mathcal{F}_{c})$ is
\begin{equation}
\eta_y^2= \frac{a_0(T_{c,0}-T)+g_4\epsilon_{B_{1g}}}{2b_1}=\frac{a_0(T_{1}-T)}{2b_1}.
\label{eqn:etay}
\end{equation}
Further, the specific heat jump at this transition, calculated by using the above $\eta_y^2$, is
\begin{equation}
\bigg(\frac{\Delta C}{T}\bigg)_1=\frac{a_0^2}{2b_1}.
\end{equation}

Below $T_1$, the system undergoes TRSB transition when $\eta_x$ condenses. Naively, one might expect this to occur at $T_{c,0}-g_4\epsilon_{B_{1g}}/a_0$, found by setting the quadratic coefficient of $\eta_x$ to zero. The condensation of $\eta_y$ prevents this, however, through the $\eta_x^2\eta_y^2$ terms in $\mathcal{F}_{op}$. If the coefficient of this term is zero ($2b_1+b_3-b_2 = 0$), then there is no competition between $\eta_x$ and $\eta_y$, in which case the second transition does occur at $T$ \mbox{\tiny $\!\!\!\!_{TRSB}$} $=T_{c,0}-g_4\epsilon_{B_{1g}}/a_0$. 

\begin{figure*}
	\centering
	\includegraphics[width=.95\linewidth]{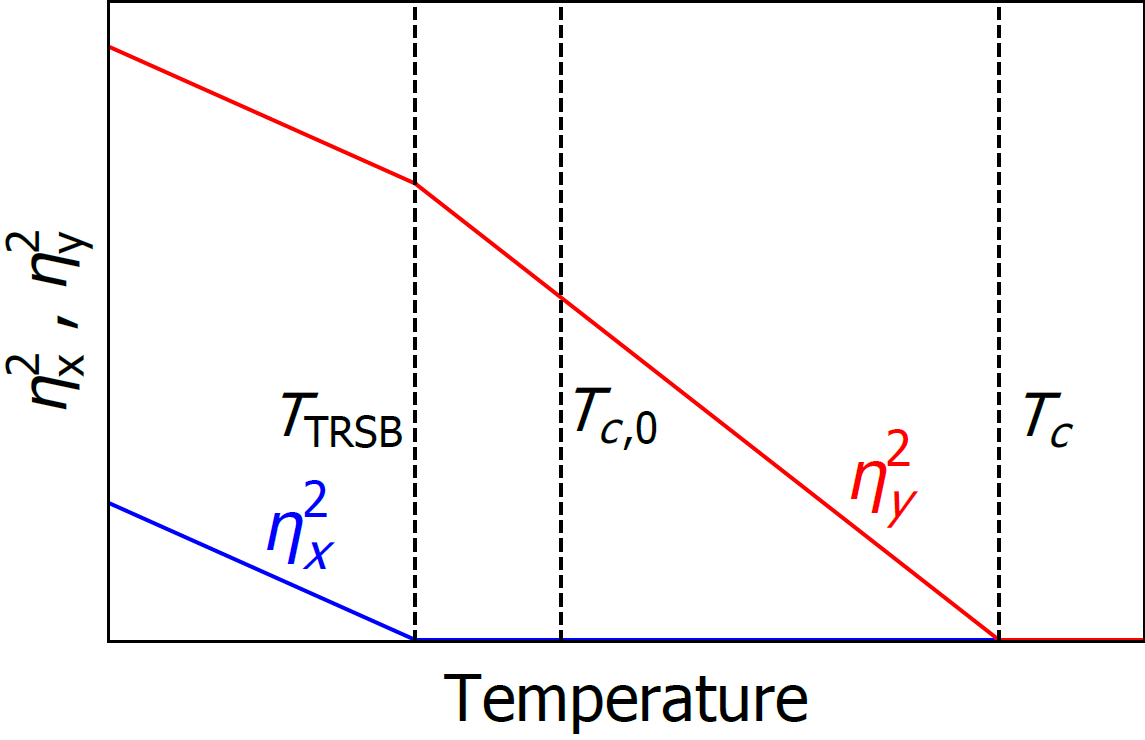}
	\caption{\textbf{Strain-induced splitting of the transition temperature $T_{c,0}$.} Under $B_{1g}$ shear strain, the two components $\eta_y$ and $\eta_x$ condense at different temperatures, T$_c$ and $T$ \mbox{\tiny $\!\!\!\!_{TRSB}$}, respectively. Above T$_c$, both the components are zero, and the sample is not superconducting. At T$_c$, the Meissner effect sets in, and finally, below $T$ \mbox{\tiny $\!\!\!\!_{TRSB}$}, the system becomes a chiral superconductor. Note that the condensation of $\eta_x$ decreases the rate at which $\eta_y$ was growing below T$_c$. Qualitatively similar behavior is expected for $B_{2g}$ shear strain, see text for details.}
	\label{fig:twoOPs}
\end{figure*}

For the more general case, when $2b_1+b_3-b_2 \neq 0$, $T_2=T$ \mbox{\tiny $\!\!\!\!_{TRSB}$} is calculated by setting the coefficient of $\eta_x^2$ to zero in the total free energy, with $\eta_y$ given by \autoref{eqn:etay}. This gives
\begin{equation}
\begin{aligned}[b]
&a_0(T_2-T_{c,0})+(2b_1+b_3-b_2)\eta_y^2+g_4\epsilon_{B_{1g}}=0\\
\implies&T_2=T_{c,0}-\frac{g_4}{a_0}\bigg(\frac{4b_1-b_2+b_3}{b_2-b_3}\bigg)\epsilon_{B_{1g}}
\end{aligned}
\end{equation}
The specific heat jump at this transition can be calculated by subtracting the jump in first transition $(\Delta C/T)_1$ from the total jump $\Delta C/T$ in the unstrained case.
\begin{equation}
\bigg(\frac{\Delta C}{T}\bigg)_2=\frac{2a_0^2}{4b_1-b_2+b_3}-\frac{a_0^2}{2b_1}=\frac{a_0^2}{2b_1}\frac{b_2-b_3}{4b_1-b_2+b_3}
\end{equation}
The ratio of the two specific heat jumps can then be related by
\begin{equation}
\bigg(\frac{\Delta C}{T}\bigg)_1\bigg/\bigg(\frac{\Delta C}{T}\bigg)_2=\frac{4b_1-b_2+b_3}{b_2-b_3}=\bigg|\frac{dT_2}{d\epsilon_{B_{1g}}}\bigg|\bigg/\bigg|\frac{dT_1}{d\epsilon_{B_{1g}}}\bigg|
\end{equation}
Below $T_2$, the order parameter $(\eta_x,\eta_y)$ can be calculated by minimizing $(\mathcal{F}_{op}+\mathcal{F}_{c})$ with respect to both $\eta_x$ and $\eta_y$. This gives 
\begin{equation}
\begin{aligned}[b]
\eta_x^2&=\frac{a_0(T_2-T)}{4b_1-b_2+b_3}\\
\eta_y^2&=\frac{a_0}{2b_1}\bigg((T_{1}-T)-\frac{2b_1-b_2+b_3}{4b_1-b_2+b_3}(T_2-T)\bigg)
\end{aligned}
\end{equation} 
It is interesting to note that the condensation of $\eta_x$ at $T_2$ decreases the rate at which $\eta_y$ was growing below $T_1$, demonstrating the competition between the two components (see \autoref{fig:twoOPs}).

The jump in the $B_{1g}$ shear modulus for chiral OP can now be expressed as,
\begin{equation}
\Delta c_{B_{1g}}=\frac{2g_4^2}{b_2-b_3}=\frac{2a_0^2}{4b_1-b_2+b_3}\cdot\frac{g_4}{a_0}\cdot\frac{g_4}{a_0}\bigg(\frac{4b_1-b_2+b_3}{b_2-b_3}\bigg)=-\frac{\Delta C}{T} \bigg|\frac{dT_1}{d\epsilon_{B_{1g}}}\bigg|\bigg|\frac{dT_2}{d\epsilon_{B_{1g}}}\bigg|
\end{equation} 

A similar derivation can be carried out for $B_{2g}$ strain. This can be performed simply by re-defining the order parameter variables as $\tilde{\eta}_x = (\eta_x+\eta_y)/\sqrt{2}$ and $\tilde{\eta}_y = (\eta_x-\eta_y)/\sqrt{2}$ and carrying out the same calculation as the $B_{1g}$ case.

The above derivation assumes that the spontaneous strains generated in the crystal below the first transition are small, such that the quartic coefficients in Landau theory are not strongly renormalized below $T_1$. We also assume that the TRS-breaking transition under finite strain is a second order phase transition; whether this holds in \sro is a question for future studies.

\subsection*{Reconciling the $c_{66}$ Discontinuity with Experiments Under Finite Strain}

It follows from our measurement of a non-zero discontinuity in $c_{66}$ that two transitions should occur under $\epsilon_{B_{2g}}=\epsilon_{xy}$ strain, each showing a specific heat jump, and \Tc must split linearly with $\epsilon_{xy}$ strain (see \autoref{eqn:Tcplus}). However, past experiments\cite{HicksScience2014,LiArxiv2019} have not observed either of these effects, and we comment on that here.

Experimental resolution of specific heat measurements under strain \cite{LiArxiv2019}, and a lack of an observed discontinuity at a lower transition ($T_{\mathrm{TRSB}} \equiv T_2$) gives a bound on the ratio of specific heat jumps at the purported transitions $1$ and $2$ as,
\begin{equation}
\bigg(\frac{\Delta C}{T}\bigg)_1\bigg/\bigg(\frac{\Delta C}{T}\bigg)_2=\bigg|\frac{dT_2}{d\epsilon_{B_{2g}}}\bigg|\bigg/\bigg|\frac{dT_1}{d\epsilon_{B_{2g}}}\bigg| \geq 20.
\end{equation}
From the jumps in the $B_{2g}$ modulus, $\Delta c_{66}\approx10^6$ Pa, and in the specific heat, $\Delta C/T=25$ mJ~mol$^{-1}$~K$^{-2}\approx450$ J~m$^{3}$~K$^{-2}$, we get
\begin{equation}
\bigg|\frac{dT_1}{d\epsilon_{B_{2g}}}\bigg|\bigg|\frac{dT_2}{d\epsilon_{B_{2g}}}\bigg|\approx 2000~ \mathrm{K}^2.
\end{equation}
From these two equations, we can estimate the shifts in the transition temperatures with strain as
\begin{equation}
\begin{aligned}[b]
&\bigg|\frac{dT_1}{d\epsilon_{B_{2g}}}\bigg|\leq 10~ \mathrm{K} = 0.1~ \mathrm{K}/\%_{\mathrm{strain}}\\
&\bigg|\frac{dT_2}{d\epsilon_{B_{2g}}}\bigg|\geq 200~ \mathrm{K} = 2~ \mathrm{K}/\%_{\mathrm{strain}}
\end{aligned}
\end{equation}

We can now compare these estimates to what has been experimentally observed. \Tc as a function of $\epsilon_{xy}$ was reported in \citet{HicksScience2014}, and the resolution on a possible cusp was ~0.1 K/\%; therefore the data of \citet{HicksScience2014} do not rule out a cusp of the magnitude predicted here. Furthermore, in recent $\mu$SR experiments under applied $B_{1g}$ strain\cite{grinenko2020split}, a modest suppression of $T$ \mbox{\tiny $\!\!\!\!_{TRSB}$ } is reported: $\sim 0.2$ K under a stress of -0.28 GPa, or a strain of $\sim-0.2\%$, for a slope of $\sim 1~ \mathrm{K}/\%$. Although measurements under $B_{2g}$ strain have not yet been reported, we note that this is comparable to the slope indicated above.

\end{document}